\documentclass[longauth]{aa}

\usepackage[varg]{txfonts}

\usepackage{siunitx}
\usepackage{float}
\usepackage[table]{xcolor}
\usepackage{hyperref}

\newcommand{\MPE}		 	{1}
\newcommand{\LIRA}		 	{2}
\newcommand{\CotedAzur}  	{3}
\newcommand{\IPAG}		 	{4}
\newcommand{\MPIA}		 	{5}
\newcommand{\UCologne}	 	{6}
\newcommand{\ESOGarching}	{7}
\newcommand{\USouthampton}	{8}
\newcommand{\KUL}	{9}
\newcommand{\FEUP}		 	{10}
\newcommand{\UBerkley}   	{11}
\newcommand{\TUM}	 	 	{12}
\newcommand{\CENTRA}	 	{13}
\newcommand{\UCD}		{14}
\newcommand{\UNAM}		{15}
\newcommand{\CRAL}		{16}
\newcommand{\KAVLI}         {17}
\newcommand{\ESOSantiago}	{18}
\newcommand{\USM}		{19}
\newcommand{\RUN}		{20}
\newcommand{\MPA}		{21}
\newcommand{\LMU}		{22}
\newcommand{\GCA}		{23}
\newcommand{\UPV}       {24}
\newcommand{\ECO}       {25}

\begin{document}

\title{GRAVITY-RESOLVE: \\A comprehensive image reconstruction tool for \\Galactic Center observations with GRAVITY}
\titlerunning{GRAVITY-RESOLVE}

\author{
K.~Abd El Dayem$^{\LIRA}$
\and N.~Aimar$^{\FEUP,\CENTRA}$
\and P.~Amaro-Seoane$^{\UPV,\MPE}$
\and B.~Aytaç$^{\MPE}$
\and A.~Berdeu$^{\ESOGarching,\LIRA}$
\and J.-P.~Berger$^{\IPAG}$
\and G.~Bourdarot$^{\MPE}$
\and W.~Brandner$^{\MPIA}$
\and Y.~Cao$^{\MPE}$
\and C.~Correia$^{\FEUP,\CENTRA}$
\and R.~Davies$^{\MPE}$
\and D.~Defr{\`e}re$^{\KUL}$
\and A.~Drescher$^{\IPAG}$
\and F.~Eisenhauer$^{\MPE,\TUM}$
\and T. A. Enßlin$^{\MPA,\LMU,\GCA}$
\and M.~Fabricius$^{\MPE}$
\and H.~Feuchtgruber$^{\MPE}$
\and S.~Flesch$^{\MPE}$
\and N.M.~F{\"o}rster~Schreiber$^{\MPE}$
\and A.~Foschi$^{\LIRA}$
\and Q.~Fournier$^{\MPE}$
\and P.~Garcia$^{\FEUP,\CENTRA}$
\and R.~Garcia~Lopez$^{\UCD}$
\and R.~Genzel$^{\MPE,\UBerkley}$
\and S.~Gillessen$^{\MPE}$ 
\and F.~Gont{\'e}$^{\ESOGarching}$
\and X.~Haubois$^{\ESOSantiago}$
\and S.F.~H{\"o}nig$^{\USouthampton}$ 
\and M.~Houll{\'e}$^{\IPAG}$
\and S.~Joharle$^\MPE$
\and J.~Kammerer$^\ESOGarching$
\and A.~Kaufer$^{\ESOSantiago}$
\and P.~Kervella$^{\LIRA}$
\and J.~Knollmüller$^{\TUM,\RUN}$
\and L.~Kreidberg$^{\MPIA}$
\and L.~Labadie$^\UCologne$
\and R.~Laugier$^{\KUL}$
\and S.~Lacour$^{\LIRA, \ESOGarching}$
\and O.~Lai$^{\CotedAzur}$
\and J.-B.~Le~Bouquin$^{\IPAG}$
\and J.~Leftley$^{\USouthampton}$
\and R.~Li$^{\MPE}$
\and B.~Lopez$^{\CotedAzur}$
\and D.~Lutz$^{\MPE}$
\and F.~Mang$^{\MPE,\TUM}$\thanks{Corresponding authors: Felix Mang, e-mail: \texttt{fmang@mpe.mpg.de}}
\and A.~M{\'e}rand$^{\ESOGarching}$
\and F.~Millour$^{\CotedAzur}$
\and M.~Montarg{\`e}s$^{\LIRA}$       
\and N.~Moruj{\~a}o$^{\FEUP,\CENTRA}$
\and E.~Nehiba$^{\MPE}$
\and H.~Nowacki${^\CotedAzur}$
\and M.~Nowak${^\LIRA}$
\and J.~Osorno${^\LIRA}$
\and T.~Ott$^{\MPE}$      
\and S.~Pappert$^{\MPE}$
\and T.~Paumard$^{\LIRA}$
\and C.~Paladini$^{\ESOSantiago}$
\and K.~Perraut${^\IPAG}$
\and G.~Perrin$^{\LIRA}$
\and R.~Petrov$^{\CotedAzur}$
\and P.O.~Petrucci$^{\IPAG}$
\and N.~Pourr{\'e}$^{\IPAG}$        
\and S.~Rabien$^{\MPE}$
\and D.C.~Ribeiro$^{\MPE}$ 
\and S.~Robbe-Dubois$^{\CotedAzur}$
\and M.~Sadun~Bordoni$^{\MPE}$
\and J.~S{\'a}nchez~Berm{\'u}dez$^{\UNAM}$
\and J.~Sauter$^{\MPIA}$
\and J.~Scigliuto$^{\CotedAzur}$
\and J.~Shangguan$^{\KAVLI, \MPE}$
\and T.T.~Shimizu$^{\MPE}$
\and J.~Stadler$^{\USM,\ECO}$
\and F.~Soulez$^{\CRAL}$
\and C.~Straubmeier$^{\UCologne}$
\and E.~Sturm$^{\MPE}$
\and M.~Subroweit$^{\UCologne}$
\and C.~Sykes$^{\USouthampton}$
\and L.J.~Tacconi$^{\MPE}$
\and P.~Thevenet$^{\LIRA}$
\and F.~Thiel$^{\MPE}$
\and V.~Thoss$^{\MPE}$
\and I.~Urso$^{\LIRA}$
\and F.~Vincent$^{\LIRA}$
\and S.~von Fellenberg$^{\MPE}$
\and J.~Woillez$^{\ESOGarching}$     
\and G.~Zins$^{\ESOSantiago}$ 
} 

\institute{
\textit{Max Planck Institute for Extraterrestrial Physics, Giessenbachstra{\ss}e 1, 85748 Garching, Germany }
\and \textit{LIRA, Observatoire de Paris, Universit{\'e} PSL, CNRS, Sorbonne Universit{\' e}, Universit{\' e} de Paris, 5 place Jules Janssen, 92195 Meudon, France   }
\and \textit{Universit{\'e} C{\^o}te d{'}Azur, Observatoire de la C{\^o}te  d{'}Azur, CNRS, Laboratoire Lagrange, France}
\and \textit{Univ. Grenoble Alpes, CNRS, IPAG, 38000 Grenoble, France },
\and \textit{Max Planck Institute for Astronomy, K{\"o}nigstuhl 17, 69117 Heidelberg, Germany}
\and \textit{1st Institute of Physics, University of Cologne, Z{\"u}lpicher Straße 77, 50937 Cologne, Germany }
\and \textit{European Southern Observatory, Karl-Schwarzschild-Stra{\ss}e 2, 85748 Garching, Germany }
\and \textit{School of Physics \& Astronomy, University of Southampton, Southampton, SO17 1BJ, United Kingdom}
\and \textit{Institute of Astronomy, KU Leuven, Celestijnenlaan 200D, 3001, Leuven, Belgium}
\and \textit{Faculdade de Engenharia, Universidade do Porto, rua Dr. Roberto Frias, 4200-465 Porto, Portugal }
\and \textit{Departments of Physics \& Astronomy, Le Conte Hall, University of California, Berkeley, CA 94720, USA }
\and \textit{Department of Physics, Technical University of Munich, 85748 Garching, Germany }
\and \textit{CENTRA - Centro de Astrof{\' i}sica e Gravita\c{c}{\~a}o, IST, Universidade de Lisboa, 1049-001 Lisboa, Portugal }
\and \textit{School of Physics, University College Dublin, Belfield, Dublin 4, Ireland}
\and \textit{Instituto de Astronom{\'i}a, National Autonomous University of Mexico, Mexico City, Mexico}
\and \textit{Univ. Lyon, Univ. Lyon 1, ENS de Lyon, CNRS, Centre de Recherche Astrophysique de Lyon UMR5574, F-69230, Saint Genis-Laval, France},
\and \textit{Kavli Institute for Astronomy and Astrophysics, Peking University, Beijing, China }
\and \textit{European Southern Observatory, Casilla 19001, Santiago 19, Chile }
\and \textit{University Observatory, Faculty of Physics, Ludwig-Maximilians-Universität, Munich, Germany}
\and \textit{Department of Astrophysics, Radboud University, Heyendaalseweg 135, Nijmegen, 6525 AJ, The Netherlands}
\and \textit{Max Planck Institute for Astrophysics, Karl-Schwarzschild-Str. 1, 85748 Garching, Germany}
\and \textit{Faculty of Physics, Ludwig-Maximilians-Universität (LMU), Geschwister-Scholl-Platz 1, 80539 Munich, Germany}
\and \textit{German Center for Astrophysics, Postplatz 1, 02826 Görlitz, Germany}
\and \textit{Universitat Politècnica de València, Valencia, Spain}
\and \textit{Excellence Cluster ORIGINS, Boltzmannstr. 2, 85748 Garching, Germany}
}

\date{Received date /
Accepted date }

\abstract
{Stars orbiting the supermassive black hole SgrA* in the Galactic Center serve to probe effects predicted by general relativity in the weak field regime. One of the effects, the Lense-Thirring precession, is a result of the black hole's angular momentum (spin), and observations of this precession could potentially constrain the spin. The precession's steep drop-off with distance makes the spin hard to measure, and until recently, no star has been found, within currently accessible magnitudes, with an orbit tight and eccentric enough to be sensitive to this effect. It is therefore interesting to increase the limiting magnitude to find faint yet undiscovered stars.}
{We aim to improve and extend the imaging capabilities of the GRAVITY beam combiner instrument of ESO's VLTI with respect to reconstruction software in the Galactic Center. We revise and develop the image reconstruction code GRAVITY-RESOLVE $\text{G}^{\text{R}}$, which has been specifically developed to reconstruct stars in the Galactic Center with GRAVITY. The main objectives are to increase the limiting magnitude in GRAVITY images, reduce its computational runtime, and allow for a joint reconstruction of multiple, spatially separate datasets/pointings on sky.}
{We present a comprehensively revised model of the Galactic Center, including a time-dependent model of the flux density of SgrA*, which invokes its measured temporal power spectrum. We enable the image reconstruction of complex visibilities, allowing us to make use of GRAVITY's full potential, reducing noise in images and referencing multiple pointings to each other. This opens the door to mosaicking in near-infrared (NIR) interferometry. The code is restructured to resemble the model's internal hierarchy, allowing for an easy swap of the underlying object model to apply it to other scientific cases.}
{We apply $\text{G}^{\text{R}}$ to recent GRAVITY data, featuring the implemented GRAVITY-faint mode as part of GRAVITY+, the upgrade of GRAVITY. We reconstruct visibility phases and amplitudes and present the first GRAVITY image reconstructed with $\text{G}^{\text{R}}$ from these quantities. We demonstrate NIR mosaicking with $\text{G}^{\text{R}}$ by jointly reconstructing nine separately acquired datasets of 2025 into a single image on a scale of $\simeq\SI{3390}{AU}\times\SI{3390}{AU}$, corresponding to $\SI{409.6}{mas}\times\SI{409.6}{mas}$. With dedicated injection tests of point sources into GRAVITY data, we estimate the limiting magnitude to be $m_{\text{K}}\ \simeq\ 20.0$ within a radius of $\SI{25}{mas}$ around the field center when reconstructing from closure phases and visibility amplitudes. We report on the discovery of two new faint stars in the Galactic Center: S303 and S304. Finally, we find a decrease in computational runtime by a factor of at least $\sim10$.}
{With $\text{G}^{\text{R}}$, we demonstrate GRAVITY's capabilities for NIR imaging. We show that $\text{G}^{\text{R}}$ is a fast, precise, and deep image reconstruction tool for Galactic Center observations, enabling the discovery of even more faint stars.}


\maketitle
\nolinenumbers

\section{Introduction}
The Galactic Center (GC) is a unique environment to test general relativity in the gravitational potential of a supermassive black hole (SMBH) due to its proximity compared to other galactic nuclei \citep{genzel_galactic_2010}. The GC's SMBH SgrA* is enshrouded in an isotropic 
 cluster of S-stars on eccentric orbits \citep{gillessen_update_2017}. Astrometry and radial velocity measurements of the orbits of the S-stars yielded a precise measurement of the distance and mass to/of SgrA* \citep{gravity_collaboration_improving_2024, abuter_mass_2022, abuter_geometric_2019, do_relativistic_2019, gillessen_update_2017, boehle_improved_2016} and have delivered an upper bound of the enclosed mass of down to ${\sim}1200\ M_{\odot}$ within the orbit of S2 at $1\sigma$ confidence \citep{gravity_collaboration_improving_2024, abuter_mass_2022}. Moreover, the S-stars can also serve as test particles to probe effects beyond Newtonian gravity \citep{zucker_probing_2006,weinberg_stellar_2005, rubilar_periastron_2001,jaroszynski_relativistic_1998}.
One of the brightest members of this cluster, S2, is particularly prone to effects of General Relativity due to its comparably eccentric orbit and short semi-major axis \citep{grould_general_2017}.
With S2, two post-Newtonian (PN) effects have been detected with the interferometric beam combiner instrument GRAVITY \citep{abuter_first_2017} of ESO's VLTI \citep{scholler_very_2007}. By closely monitoring the proper motion of S2 during its pericenter passage with GRAVITY's unmatched astrometric precision \citep{abuter_first_2017}, the Schwarzschild precession in the orbit of S2 \citep{abuter_mass_2022, abuter_detection_2020}, the gravitational redshift along with the relativistic transverse Doppler effect \citep{gravity_collaboration_detection_2018-1}, have been detected with high significance. S2 also delivered the first test of Einstein's equivalence principle in the gravitational field of a SMBH \citep{gravity_collaboration_test_2019}.\\
According to the no-hair theorem, a stationary black hole (BH) is characterized by its mass, spin, and electric charge \citep[e.g.][]{israel_event_1967}. Since the mass of SgrA* has been constrained across the past three decades with ever-increasing accuracy and precision \citep{haller_stellar_1996,eckart_observations_1996,genzel_dark_1996,eckart_stellar_1997,genzel_nature_1997,ghez_high_1998,schodel_star_2002,schodel_galactic_2003,ghez_first_2003,ghez_stellar_2005,eisenhauer_sinfoni_2005,ghez_measuring_2008, gillessen_monitoring_2009,gillessen_orbit_2009,boehle_improved_2016,gillessen_update_2017,do_relativistic_2019,abuter_geometric_2019,abuter_mass_2022}, the next step is to constrain the BH's spin. The "Eigenrotation" of a BH induces 1.5 PN order corrections to the PN expansion of the equations of motion and effectively shifts both the orbit as well as the orbital plane of circling bodies, resulting in a precession, known as Lense-Thirring precession \citep{alexander_stellar_2005, lense_uber_1918} (hereafter LT-precession).\\
The star S2 is, in theory, susceptible to this effect, but its minute LT-precession is out of reach of current facilities or would only be detectable in long-term observing campaigns after multiple revolutions of its orbit \citep{grould_general_2017}. Additionally, stars and stellar remnants could imitate this precession by means of vector resonant relaxation \citep{merritt_testing_2010, rauch_resonant_1996}. Moreover, the LT-precession drops as $\propto r^{-3}$ with distance $r$ to the central massive object \citep{will_testing_2008}. To measure the spin, hence, requires the discovery of stars that are orbiting closer to SgrA*. To find new, yet undiscovered, faint stars in the GC, which offer potentially closer pericenter distances, image reconstruction of interferometric data is the tool of choice.\\
Interferometric imaging, in both radio and near-infrared (NIR) wavelength regimes, is a powerful approach as the limitations of single-aperture imaging are overcome. The size of a synthesized aperture array is only limited by its infrastructure, allowing to achieve almost arbitrarily high spatial resolutions.\\
The multitude of science cases to which interferometric imaging can be applied led to an entire plethora of image reconstruction methods that have emerged over the past decades. Pioneered by developments in radio interferometry \citep{abuter_detection_2021,greisen_aips_2003,hogbom_aperture_1974} and fostered by an imaging contest \citep[e.g.][]{lawson_interferometry_2004}, a number of reconstruction codes in NIR interferometry have been introduced. Most of optical/NIR interferometry reconstruction software is general purpose, i.e. can be applied to a wide variety of objects and comprise ORGANIC \citep{claes_neural_2020}, IRBis \citep{hofmann_image_2014}, SQUEEZE \citep{baron_novel_2010}, MIRA \citep{thiebaut_mira_2008}, MACIM \citep{ireland_monte-carlo_2006}, WISARD \citep{le_besnerais_advanced_2008}, and BSMEM \citep{buscher_direct_1994}, while other tools like SPARCO \citep{kluska_sparco_2014}, and ROTIR \citep{martinez_dynamical_2021}, SURFING \citep{roettenbacher_no_2016}, are focused on reconstructing Young Stellar Objects and stellar surfaces, respectively.\\
In this work, we present a comprehensive revision, update, and extension of the image reconstruction method GRAVITY-RESOLVE\footnotemark $\text{G}^{\text{R}}$ \citep{gravity_collaboration_deep_2022} and give a new upper limit of the limiting magnitude of GRAVITY in the Galactic Center.
In Section \ref{section_methods} we outline the revised model of the Galactic Center implemented in $\text{G}^{\text{R}}$. Furthermore, we describe a method for reconstructing a mosaic of spatially separated datasets. In Section \ref{section_data}, we give an overview of the GRAVITY data, which is used for reconstructing complex visibilities, the demonstration of the mosaicking approach, and a performance test of $\text{G}^{\text{R}}$ in Section \ref{section_results}. Finally, in Section \ref{section_discussion} we discuss the results and give concluding remarks in Section \ref{section_conclusion}.
\footnotetext{GRAVITY-RESOLVE is available at:\\ \url{https://gitlab.mpcdf.mpg.de/mpe/ir/gravity-resolve}}
\section{Methods}\label{section_methods}
In interferometry, the coherence of light between two telescopes is quantified by the complex visibility. The van Cittert-Zernike theorem \citep{van_cittert_wahrscheinliche_1934, zernike_concept_1938} relates the complex visibility $V$ (hereafter just visibility) to the apparent flux density distribution of the object $F\left(\boldsymbol{s}\right)$ via a Fourier Transform,
\begin{equation}\label{eq_van_cittert-zernike}V_{\text{j,k}}\left(\boldsymbol{u}\right)=V_{\text{j,k}}\left(\frac{\boldsymbol{B}_{\text{j,k}}}{\lambda}\right)=\frac{\int\,d\boldsymbol{s}\ F\left(\boldsymbol{s}\right)\ e^{-2\pi i\boldsymbol{s}\frac{\boldsymbol{B}_{\text{j,k}}}{\lambda}}}{\int\,d\boldsymbol{s}\ F\left(\boldsymbol{s}\right)},
\end{equation}
with $\boldsymbol{s}$ the position vector on sky. A pair of telescopes $\text{j}$ and $\text{k}$ form a baseline $\boldsymbol{B}_{\text{j,k}}$, that is, their projected separation vector on sky. The baseline and the observed wavelength $\lambda$ determine the spatial frequency $\boldsymbol{u}$ at which the visibility is sampled.\\
With a fixed number of stationary telescopes, altering the observed spatial frequencies is limited to the number of spectral channels and observing time available, since baselines change length and orientation as Earth rotates with respect to the object. The limited number of telescopes in NIR interferometry in comparison to radio interferometry results in particularly sparse datasets.\\
In the case of GRAVITY, Galactic Center observations are carried out with the four stationary Unit Telescopes (UTs) of the VLTI in low spectral resolution \citep{abuter_first_2017}, yielding in principle 1680 complex datapoints, or visibilities, for an average dataset of $10$ exposures.\\
With $\text{G}^{\text{R}}$, we aim to reconstruct an image that has at least one order of magnitude more degrees of freedom than there are datapoints available, which impedes the blunt inversion of the response equation (Eq. \ref{eq_van_cittert-zernike}) to obtain an image, posing an underdetermined problem.\\
The likelihood of this problem does not uniquely constrain the model, and prior information is needed to tame degeneracies. Bayes' Theorem \citep{bayes_lii_1763} describes exactly this setting, in which a likelihood $\mathcal{P}\left(d|F\right)$ with data $d\coloneqq V_{\text{obs}}$, given an object model $F$, is informed by a prior $\mathcal{P}\left(F\right)$ of the model, to obtain the posterior $\mathcal{P}\left(F|d\right)$,
\begin{equation}\label{eq_bayes_theorem}
    \mathcal{P}\left(F|d\right)=\frac{\mathcal{P}\left(d|F\right)\mathcal{P}\left(F\right)}{\mathcal{P}\left(d\right)}\propto\mathcal{P}\left(d|F\right)\mathcal{P}\left(F\right),
\end{equation}
where $\mathcal{P}\left(d\right)$ is the evidence and $\mathcal{P}$ denotes a probability density function (PDF).\\
We apply the theoretical framework of information field theory (IFT) \citep{enslin_information_2022,enslin_information_2019,enslin_physics_2019} to develop a hierarchical model of the Galactic Center and the instrument response of GRAVITY. A numerical implementation of IFT in \texttt{python}, NIFTy \citep{edenhofer_re-envisioning_2024,arras_nifty5_2019,steininger_nifty_2019,selig_nifty_2013} is used to formulate the model in code and to deploy one of the inference methods \citep{knollmuller_metric_2020} that are available in NIFTy.\\
Since the evidence $\mathcal{P}\left(d\right)$ is hard to compute and does not depend on the object model $F$ \citep{knollmuller_metric_2020,gravity_collaboration_deep_2022}, we renounce a normalized posterior. Instead, the quantity that is explored is the information Hamiltonian of the posterior $\mathcal{H}\left(F|d\right)$ \citep{enslin_information_2019,knollmuller_metric_2020},
\begin{equation}\label{eq_information_hamiltonian}
    \mathcal{H}\left(F|d\right)\ = \ -\text{ln}\left(\mathcal{P}\left(F|d\right)\right)\ \hat{=}\ \mathcal{H}\left(d|F\right)+\mathcal{H}\left(F\right),
\end{equation}
which is equal to the sum of the likelihood and prior information Hamiltonian, up to the constant evidence term, symbolized by the circumflex $\left(\hat{=}\right)$.
\subsection{Modeling the Galactic Center}\label{subsection_gc_model}
The model of the Galactic Center in $\text{G}^{\text{R}}$ consists of three parts, that is, the central SMBH SgrA*, known bright stars \citep[$m_{\text{K}} < 18.9$][]{abuter_detection_2021} within the field of view (FOV), and a spatially extended background that accounts for any faint, yet unknown flux contributions, which ideally amount to undiscovered sources. $\text{G}^{\text{R}}$ is designed to reconstruct stars that appear to GRAVITY as unresolved point sources.\\
SgrA* is the most intricate component of the model due to its strong variability in both quiescent and flaring phases \citep{dodds-eden_two_2011}, demanding a model of flux density over time. This is the only time-dependent aspect of the object model, besides instrumental calibration parameters. As the variability of SgrA* is already incorporated in $\text{G}^{\text{R}}$, an exposure-by-exposure reconstruction as with AIPS's CLEAN \citep{abuter_detection_2021,greisen_aips_2003,hogbom_aperture_1974} is avoided.\\
The lightcurve of SgrA* exhibits a temporal correlation, which we take into account in our model by invoking the measured power spectrum \citep{do_near-infrared_2009, dodds-eden_evidence_2009, witzel_variability_2018, witzel_rapid_2021} in the K-band \citep{yusef-zadeh_nonstop_2025}. Moreover, we model the flux density distribution of SgrA* as an inverse gamma distribution, accounting for the fact that it follows a log-normal distribution, which is extended by a flattened tail to higher fluxes to account for flaring episodes \citep{abuter_flux_2020}. The spectral energy distribution of SgrA* is described by a power-law \citep{yusef-zadeh_nonstop_2025, witzel_variability_2018}, with a dependence of the spectral index on flux density \citep{yusef-zadeh_nonstop_2025}. We model the spectral index to be normally distributed and account for its variability with an accordingly large standard deviation. Finally, the position of SgrA* $\boldsymbol{s}_{\text{SgrA}}$ on sky is assumed to follow also a normal distribution both in RA and Dec, reflecting the statistical offset of the single-mode fibers in the GRAVITY beam combiner instrument \citep{abuter_first_2017}.\\
In $\text{G}^{\text{R}}$, bright sources in GRAVITY's FOV are parameterized by their position, flux density, and spectral index. The position prior follows a normal distribution with the same width as with SgrA*, centered on predicted positions, which are based on our relativistic orbit models. The flux density is modeled as a log-normal distribution, with the favorable property of being strictly positive. The statistics for this distribution stem from photometric samples acquired over decades \citep{gillessen_update_2017}. Finally, we assume the spectral index to follow a normal distribution centered on the power of the Rayleigh-Jeans law \citep{buchholz_composition_2009} and choose a generous width, such that this spectral index also holds for faint, i.e., cool main-sequence stars in the GC.
The model for faint sources is a continuous field of pixels, with every pixel exhibiting the same flux density distribution based on the K-band luminosity function (KLF) in the GC. The slope of this distribution is augmented by an additional slope parameter that allows for the inference of systematic background flux density.\\
Taken together, the object model $F\left(\boldsymbol{s},\lambda,t,p\right)$ in $\text{G}^{\text{R}}$ extends across the spatial, spectral, temporal domain, and polarizations (cf. \ref{subsection_overall_object_model}). The object model of the Galactic Center is implemented in $\text{G}^{\text{R}}$ as a NIFTy operator $\mathcal{S}$ that acts on the set of latent variables $\boldsymbol{\xi}$ (cf. \ref{subsection_standardization}) to compute the corresponding flux density distribution in the GC,
\begin{equation}\label{eq_sky_op}
    \mathcal{S}\left(\boldsymbol{\xi}\right)\ \coloneqq\ F\left(\boldsymbol{s},\lambda,t,p\right).
\end{equation}
\subsection{Instrument response of GRAVITY}
We model the instrument response of GRAVITY in $\text{G}^{\text{R}}$ based on the van Cittert-Zernike theorem \citep{van_cittert_wahrscheinliche_1934, zernike_concept_1938} (Eq. \ref{eq_van_cittert-zernike}), where the object model $F\left(\boldsymbol{s}\right)$ is replaced by the model $F\left(\boldsymbol{s},\lambda,t,p\right)$ of the Galactic Center (Eq. \ref{eq_gc_model}), extending additionally in the spectral and temporal dimensions, and is dependent on the two linear polarization states.\\
The theorem is further modified to include two systematic instrumental effects of GRAVITY, that is, optical aberrations and its finite spectral bandwidth.\\
Optical aberrations in GRAVITY along individual beam paths have been measured in \cite{abuter_improved_2021} per telescope and transformed to complex interferometric lobes $L_{\text{j,k}}\left(\boldsymbol{s}\right)$ or \textit{aberration maps} per baseline in the image plane (see Appendix \ref{appendix_response} for details). In practice, residual time-dependent phase variations due to atmospheric turbulence and/or instrumental systematics, that is, tip-tilt jitter of the adaptive optics (AO) system, blur the aberration maps $\left\langle L_{\text{j,k}}\left(\boldsymbol{s}\right)\right\rangle_{\text{turb}}$ \citep{abuter_improved_2021}. In the instrument response of GRAVITY, the aberration maps are multiplied by $F\left(\boldsymbol{s},\lambda,t,p\right)$, before the Fourier transform is applied (see Appendix \ref{appendix_response}). In the picture of point sources in the GC, the amplitude of the aberration maps dims their flux density, and the phase affects the optical path difference (OPD) in their visibility. Hence, to every point source in the FOV, a correction to its flux density and position is effectively applied.\\
The finite width of spectral channels in GRAVITY leads to bandwidth smearing \citep{gravity_collaboration_deep_2022,abuter_detection_2021}, which has an increasingly distorting effect in the image plane towards the edge of the FOV. This is accounted for by dividing every spectral channel in GRAVITY into $N_{\text{sub}}$ spectral sub-channels and evaluating the respective instrument responses. Every response is weighted by the spectral bandpass in GRAVITY $P_{\lambda}\left(\lambda_{\text{sub}}\right)\ \coloneqq\ P_{\lambda\ |\ \text{i,j}}\left(\lambda_{\text{sub}},p\right)$ (Fig. \ref{fig_GRAVITY_bandpass}), which is dependent on baseline, spectral sub-channel and polarization state, and subsequently integrated. With this numerical approach, we refine the top-hat bandpass as used in \cite{gravity_collaboration_deep_2022}.\\
Moreover, we extend the self-calibration approach, applied to visibility amplitudes in \cite{gravity_collaboration_deep_2022}, to visibility phases. In the self-calibration of visibility amplitudes (Eq. \ref{eq_visamp_selfcal}), a correction factor $C_{\rho\ |\ \text{j,k}}\left(t_{\text{m}}\right)$ per baseline and exposure, that is the instance in time $t_{\text{m}}$ it was taken, is applied to the visibility amplitudes. Equivalently, we introduce a self-calibration for visibility phases (Eq. \ref{eq_visphi_selfcal}) where we apply a correction factor in OPD $C_{\phi\ |\ \text{j}}\left(t_{\text{m}}\right)$ per telescope and exposure. This heuristic model aims at absorbing time-dependent, statistical instrumental effects, lifting minor inconsistencies in the data, and ultimately leading to potentially deeper images.\\
The overall instrument response of GRAVITY (Eq. \ref{eq_response}) along with the self-calibration is implemented as a NIFTy operator in $\text{G}^{\text{R}}$. The response operator $\mathcal{R}$ acts on the object model $F$ to compute the corresponding self-calibrated model visibility samples $V_{\text{j,k}}^{\text{cal}}\left(\boldsymbol{u}\right)$,
\begin{equation}\label{eq_response_op}
    \mathcal{R}\left(F\left(\boldsymbol{s},\lambda,t,p\right)\right)\ \coloneqq\ V_{\text{j,k}}^{\text{cal}}\left(\boldsymbol{u}\right).
\end{equation}
In $\text{G}^{\text{R}}$, the response operator $\mathcal{R}$ is chained with the object model operator $\mathcal{S}$ (Eq. \ref{eq_sky_op}), forming a deeply hierarchical model to compute model visibility samples directly from the set of latent variables (cf. \ref{subsection_standardization}),
\begin{equation}
    \left(\mathcal{R}\circ\mathcal{S}\right)\left(\boldsymbol{\xi}\right)\ =\ V_{\text{j,k}}^{\text{cal}}\left(\boldsymbol{u}\right)
\end{equation}
A schematic overview of the hierarchical model in $\text{G}^{\text{R}}$ is given in Fig. \ref{fig_hierarchical_model}.
\subsection{Mosaicking}\label{subsection_mosaicing}
\begin{figure*}
    \centering
    \includegraphics[width=\linewidth]{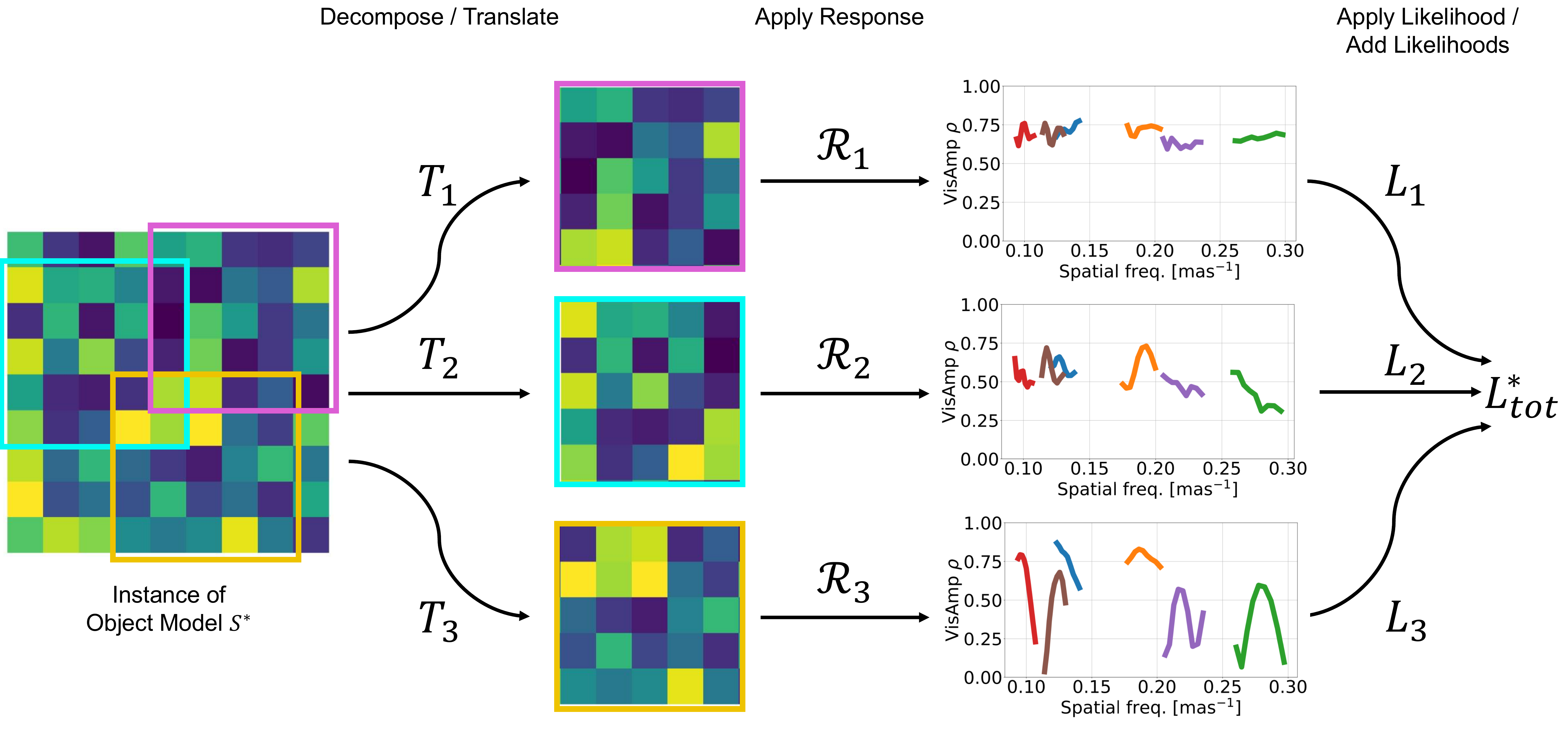}
    \caption{Conceptual overview of the mosaic forward model implemented in $\text{G}^{\text{R}}$. An instance of the object model $\mathcal{S}^{*}$ is shifted to match the center of the respective pointing $\boldsymbol{\sigma}_{n}$ and the instrument response of GRAVITY is applied to it. The likelihood is computed for the response of each pointing, and the corresponding likelihoods are subsequently summed. For details, see subsection \ref{subsection_mosaicing}.}
    \label{fig_Mosaic_model}
\end{figure*}

Mosaicking describes the \textit{joint} reconstruction of multiple, spatially separated pointings on sky into a single image \citep{sault_approach_1996}. Routinely applied in radio interferometry (see e.g. \citep{dunlop_deep_2017}), it stems from a workaround to sample shorter (or even zero) spatial frequencies to alleviate dominant defects in the synthesized beam \citep{ekers_short_1979}, equivalent to the dirty beam in NIR interferometry. Mosaicing resembles a conceptual combination of single-dish imaging, where an individual radio dish is moved to sample a large emission structure, and interferometric imaging \citep{cornwell_radio-interferometric_1988}.\\
We present a first method to achieve mosaicking in NIR interferometry. This is made possible by GRAVITY's capability to provide visibility phases \citep{abuter_first_2017}, which provide an absolute reference between individual pointings.\\
The working principle of the implemented mosaicking method is explained in Fig. \ref{fig_Mosaic_model}.\\
An instance of the object model $F^{*}\left(\boldsymbol{s}\right)$, that is, a possible realization of the flux density distribution, is first decomposed into separate regions corresponding to the positions of pointings on sky:
\begin{enumerate}
    \item We first shift the FOV by applying a two-dimensional inverse Fourier transform to the instance of the object model $F^{*}\left(\boldsymbol{s}\right)$. We then multiply a phase term, which corresponds to a shift of the FOV by the vector $\boldsymbol{\sigma}_{\text{n}}$ for pointing $\text{n}$, and use the positions of pixels in the frequency domain as spatial frequencies $\boldsymbol{k}$. Another Fourier transform yields the shifted object model (Eq. \ref{eq_decomposition_op}).
    \item To avoid wrapping of the shifted object model due to the periodic behaviour of the Fourier Transform, we apply a mask $\mathcal{M}$ that pads the corresponding area to zeros.
\end{enumerate}
Both steps are implemented in $\text{G}^{\text{R}}$ as linear NIFTy operators and when chained constitute the \textit{decomposition} operator $\mathcal{T}_{\text{n}}$ for pointing $\text{n}$,
\begin{equation}\label{eq_decomposition_op}
    \mathcal{T}_{\text{n}}\ =\ \mathcal{M}\circ\mathcal{F}\left[e^{2\pi i\boldsymbol{k}\cdot\boldsymbol{\sigma}_{n}}\mathcal{F}^{-1}\left[F\left(\boldsymbol{s}\right)\right]\right].
\end{equation}
Finally, we incorporate the decomposition operator $\mathcal{T}_{\text{n}}$ into the hierarchical model of $\text{G}^{\text{R}}$ to obtain a response operator $\mathcal{R}_{\text{n}}$ for every pointing, which corresponds to the application of the response in Fig. \ref{fig_Mosaic_model},
\begin{equation}
    \mathcal{R}_{\text{n}}\ \coloneqq\ \mathcal{R}\circ\mathcal{T}_{\text{n}}\circ\mathcal{S}.
\end{equation}
\subsection{Noise statistics of GRAVITY/Likelihood}\label{subsection_likelihood}
The data of GRAVITY can be modeled by an instrument response including systematics (optical aberrations/bandwidth smearing) with an additional statistical noise component \citep{enslin_information_2022, enslin_information_2019},
\begin{equation}
    d\ =\ \mathcal{R}\left(F\left(\boldsymbol{s},\lambda,t,p\right)\right)+n.
\end{equation}
The likelihood $\mathcal{P}\left(d|F\right)$ is defined as the marginal PDF of the data $d$ \citep{enslin_information_2022, enslin_information_2019}, by integrating $\mathcal{P}\left(d|n\right)\ =\ \delta\left(d-\mathcal{R}\left(F\right)-n\right)$ over the noise statistics of GRAVITY $\mathcal{P}\left(n|F\right)$,
\begin{equation}
    \mathcal{P}\left(d|F\right)\ =\ \int\,dn\ \delta\left(d-\mathcal{R}\left(F\right)-n\right)\mathcal{P}\left(n|F\right).
\end{equation}
The GRAVITY data reduction pipeline \citep{lapeyrere_gravity_2014} provides the visibility samples as integrations over individual detector integration times (DITs), which are long enough that we can assume, along the line of the central limit theorem, that the noise statistics are normally distributed. Moreover, the pipeline yields data in the OIFITS format \citep{pauls_data_2005,duvert_oifits_2017} that provides the first and second moments of $\mathcal{P}\left(n|F\right)$, constraining the picture of the noise distribution again to that of a normal distribution,
\begin{equation}
    \mathcal{P}\left(d|F\right)\ \sim\ \mathcal{N}\left(d-\mathcal{R}\left(F\right)|\ 0, \Sigma\right).
\end{equation}
The pipeline neglects cross-correlations, effectively reducing the noise covariance matrix to $\Sigma\ \coloneqq\ \sigma_{\text{i,j}}^{2}\mathbf{1}$.\\
The information Hamiltonian (Eq. \ref{eq_information_hamiltonian}) of the likelihood reads
\begin{equation}
    \mathcal{H}\left(d|F\right)\ \hat{=}\ \frac{1}{2}\left(d-\mathcal{R}\left(F\right)\right)^{\dagger}\Sigma^{-1}\left(d-\mathcal{R}\left(F\right)\right).
\end{equation}
For the mosaicking model, this is generalized to a summation over individual pointings,
\begin{flalign}
    & \mathcal{H}\left(d|F\right)\ \hat{=}\ \sum_{n=1}^{N_{\text{pointings}}}\frac{1}{2}\left(d_{\text{n}}-\mathcal{R}_{\text{n}}\left(F\right)\right)^{\dagger}\Sigma_{\text{n}}^{-1}\left(d_{\text{n}}-\mathcal{R}_{\text{n}}\left(F\right)\right),
\end{flalign}
bearing a strong similarity to the $\chi^{2}$ described for mosaicking in \cite{cornwell_radio-interferometric_1993}.
\subsection{Inference}\label{subsection_inference}
The posterior of the inference problem is given within the formalism of the information hamiltonian (Eq. \ref{eq_information_hamiltonian}) as the sum of the likelihood term $\mathcal{H}\left(d|F\right)$ (subsection \ref{subsection_likelihood}), and the corresponding prior term $\mathcal{H}\left(F\right)$,
\begin{equation}
    \mathcal{H}\left(F|d\right)\ \hat{=}\ \mathcal{H}\left(d|F\right)+\mathcal{H}\left(F\right).
\end{equation}
Exploring the posterior $\mathcal{H}\left(F|d\right)$ to find the best possible image $F$, given data $d$ and prior information, is a non-trivial task given the high number of parameters to be inferred and the sparsity of the data.\\
Due to the non-conjugate nature of the prior, an analytic posterior is not accessible, and we have to rely on an approximate solution \citep{knollmuller_metric_2020}. We apply Metric Gaussian Variational Inference (MGVI) \citep{knollmuller_metric_2020}, a Bayesian inference method, provided as part of the NIFTy package \citep{edenhofer_re-envisioning_2024,arras_nifty5_2019,steininger_nifty_2019,selig_nifty_2013}.\\
MGVI is well-suited for image reconstruction as it is capable of exploring extremely high-dimensional posteriors and provides accurate results in a manageable amount of time. It scales linearly in time and memory with the number of parameters, while posing only a few demands on the underlying inference problem \citep{knollmuller_metric_2020}.\\
Besides continuously distributed parameters, independently sampled data, and an adequately Gaussian posterior, MGVI demands to work on a standardized coordinate system \citep{knollmuller_metric_2020}. All parameters follow a standard normal distribution in this coordinate system, and a standardization function $f\left(\xi\right)$ maps them to their corresponding prior distribution (see Appendix \ref{subsection_standardization}) \citep{knollmuller_metric_2020, kingma_auto-encoding_2022, devroye_sample-based_1986}. All parameters defined in subsection \ref{subsection_gc_model} that follow a custom prior are based on their standardized coordinates $\xi$. This flattens the hierarchical model in $\text{G}^{\text{R}}$, and the posterior is solely dependent on the standardized coordinates,
\begin{equation}
        \mathcal{H}\left(f\left(\xi\right)|d\right)\ \hat{=}\ \mathcal{H}\left(d|f\left(\xi\right)\right)+\frac{1}{2}\xi^{\dagger}\mathbf{1}\xi.
\end{equation}
The key concept of MGVI \citep{knollmuller_metric_2020} is to approximate the true posterior $\mathcal{P}\left(\xi|d\right)$ (or its information hamiltonian $\mathcal{H}\left(f\left(\xi\right)|d\right)$), by a parametric family of PDFs, that is, it is approximated by a multivariate normal distribution $\mathcal{Q}_{\varphi}\left(\xi\right)\ =\ \mathcal{N}\left(\bar{\xi},\Theta\left(\bar{\xi}\right)\right)$ with variational paramters $\varphi$ in an iterative fashion:
\begin{enumerate}
    \item At an initial position $\bar{\xi}^{\left(0\right)}$ for the mean of the approximate posterior $\mathcal{Q}_{\varphi}\left(\xi\right)$, the local curvature of the true posterior $\mathcal{P}\left(\xi|d\right)$ is approximated by the covariance \citep{knollmuller_metric_2020},
    \begin{equation}\label{eq_approx_cov}
        \Theta^{-1}\ \approx\ \left<\frac{\partial\mathcal{H}\left(d,\xi\right)}{\partial\xi}\frac{\partial\mathcal{H}\left(d,\xi\right)}{\partial\xi^{\dagger}}\right>_{\mathcal{N}\left(\xi|\bar{\xi},\Theta\right)}.
    \end{equation}
    \item The current position is updated by maximizing the overlap between the approximate and true posterior \citep{knollmuller_metric_2020}. This is achieved by minimizing the Kullback-Leibler divergence \citep{kullback_information_1951} with respect to $\xi$,
    \begin{equation}
        \mathcal{D}_{\text{KL}}\left(\mathcal{Q}_{\varphi}\left(\xi\right)||\mathcal{P}\left(\xi|d\right)\right)\ =\ \int\,d\xi\ \mathcal{Q}_{\varphi}\left(\xi\right)\ \text{ln}\frac{\mathcal{Q}_{\varphi}\left(\xi\right)}{\mathcal{P}\left(\xi|d\right)}.
    \end{equation}
\end{enumerate}
These steps are repeated until convergence is reached. The approximate posterior $\mathcal{Q}_{\varphi}\left(\xi\right)$ is then returned as a set of samples from which the final image is computed.
\section{Data}\label{section_data}
We demonstrate the capabilities of $\text{G}^{\text{R}}$ on 11 GRAVITY Galactic Center datasets acquired in 2021 and 2025. These span one single epoch observation for a performance test of $\text{G}^{\text{R}}$, one multi-epoch observation for mosaicking, and one recent dataset featuring the latest GRAVITY+ implementation, the 'extreme AO' \citep{bourdarot_gravity_2024, collaboration_first_2025}.
\subsection{GRAVITY+ observations}\label{subsection_gpao_data}
After successful integration of the GRAVITY+ adaptive optics (GPAO) system \citep{bourdarot_gravity_2024, collaboration_first_2025} in October 2024, the first GPAO dataset during GTO (Guaranteed Time Observations) was acquired on 2025-03-15. Since the Laser Guide Stars (LGS) were not yet in place, this dataset profited mostly from the previous GRAVITY+ implementation, GRAVITY-faint \citep{widmann_gravity_2022}. In total, we recorded 14 exposures centered on SgrA*, one R2 exposure to calibrate visibility amplitudes, and three S4 exposures for visibility phase calibration. Each exposure is comprised of 12 individual detector integrations at $\SI{30}{s}$ integration time each, yielding $\SI{84}{min.}$ total integration time on SgrA*.\\
We applied the most recent GRAVITY data reduction pipeline \citep{lapeyrere_gravity_2014} to the raw data, subsequently applied the visibility amplitude calibration with R2, calibrated the visibility phases with one S4 exposure, and corrected for a small offset. A visual inspection of the data was carried out, leading to the selection of the seven best SgrA* exposures, reducing the total integration time to $\SI{42}{min.}$.\\
\subsection{Mosaic dataset}\label{subsection_mosaic_dataset}
We showcase the mosaicking approach implemented in $\text{G}^{\text{R}}$ by jointly reconstructing nine spatially separated GRAVITY+ datasets, acquired in August 2025. These comprise in total 88 exposures (see Table \ref{tab_mosaic_dataset} for details), amounting to $\SI{528}{min.}$ total integration time. The exposures are distributed across three nights, 2025-08-06, 2025-08-07, and 2025-08-09. Besides SgrA*, there are 13 S-stars expected to be present in the FOV: S2, S24, S21, S38, S62, S60, S63, S20, S42, S9, S31, S55, and S23. A number of stars exhibit a considerable projected speed, among them S9, which is moving in the time frame of observations $\sim\SI{447}{\mu as}$. This is not a negligible amount, but should not pose a large impact on the overall image reconstruction, as S9, as well as S60 and S42, are located on the edge of their respective pointing, where the astrometry gets increasingly imprecise anyway.\\
We reduced all nine datasets consistently with the GRAVITY data reduction pipeline \citep{lapeyrere_gravity_2014}, and calibrated the visibility phases of all exposures with one carefully chosen S2 exposure of their respective night. The visibility amplitudes were calibrated with exposures on R2. All exposures have been corrected for a small offset.\\
\begin{table}
    \renewcommand{\arraystretch}{1.5}
    \centering
    \begin{tabular}{c|c|c|c}
        Pointing & Epoch & $N_{\text{exp, used}}$ & Separation $\boldsymbol{\sigma}$ [mas] \\\hline\hline
        SgrA* & 2025.602 & 10 & $(0.0, 0.0)$ \\\hline
        SgrA*-S9 & 2025.602 & 16 & $(-15.0, -15.0)$ \\\hline
        S20 & 2025.600 & 10 & $(123.473, -15.955)$ \\\hline
        Dark-S9 & 2025.606 & 8 & $(40.0, -100.0)$ \\\hline
        S55 & 2025.598 & 4 & $(-17.037, -158.354)$ \\\hline
        S2 & 2025.602 & 17 & $(-22.872, 178.460)$\footnotemark \\\hline
        S303 & 2025.600 & 8 & $(6.710, 66.468)$ \\\hline
        S301 & 2025.598 & 9 & $(-72.005, 57.737)$ \\\hline
        S38 & 2025.606 & 6 & $(-151.045, 60.039)$ \\
    \end{tabular}
    \caption{Overview on GRAVITY+ exposures used for the reconstruction of the mosaic in Fig. \ref{fig_Mosaic}. The pointing name refers to the nearest object to the GRAVITY pointing, the epoch is the average of exposure timestamps of used exposures $N_{\text{exp, used}}$. The separation vector $\boldsymbol{\sigma}$ is the position of the pointing in RA, Dec, with respect to SgrA*.}
    \label{tab_mosaic_dataset}
\end{table}
\footnotetext{Exposure weighted average, maximum separation between individual S2 pointings is $\SI{71}{\mu as}$.}
\subsection{Performance dataset}\label{subsection_performance_dataset}
We use a third dataset, which serves to compare conventional GRAVITY data to GRAVITY+ data and allows us to make a statement on the performance difference of the presented $\text{G}^{\text{R}}$ method and the previous implementation of $\text{G}^{\text{R}}$ in \cite{gravity_collaboration_deep_2022}.\\
The dataset that meets both demands is a SgrA* pointing of 2021-05-29, comprised of 22 exposures, has already been featured in \citep{gravity_collaboration_deep_2022} to demonstrate image reconstruction with closure phases and served as a basis for injection tests.\\
As with the mosaic dataset, we applied the GRAVITY data reduction pipeline \citep{lapeyrere_gravity_2014}, but calibrated visibility amplitudes with one S2 exposure and corrected for a small offset.\\
We select from this dataset the best 15 exposures, amounting to a total integration time of $\SI{117}{min.}$, since each exposure in conventional GRAVITY data consists of 32 individual detector integrations with $\SI{10}{s}$ integration time each.
\section{Results}\label{section_results}
We present three results to highlight the strengths of the revised image reconstruction method $\text{G}^{\text{R}}$ for Galactic Center observations with the GRAVITY instrument.\\
First, we demonstrate the reconstruction of an image from complex visibilities, therefore exploiting GRAVITY's phase-referencing capabilities and ultimately profiting from reduced noise. We lay out the inference scheme of $\text{G}^{\text{R}}$ and explain the refinement model following \citep{edenhofer_sparse_2022}.\\
Secondly, we present an interferometric NIR mosaic reconstructed \textit{jointly} from nine GRAVITY GC datasets, again facilitated by GRAVITY's ability to measure visibility phases.\\
Finally, we conduct a performance test of the here presented $\text{G}^{\text{R}}$ method and compare our results to \cite{gravity_collaboration_deep_2022}.
\subsection{Image reconstruction with visibility phases}\label{subsection_rec_visphi}
\begin{figure}
    \centering
    \includegraphics[width=\linewidth]{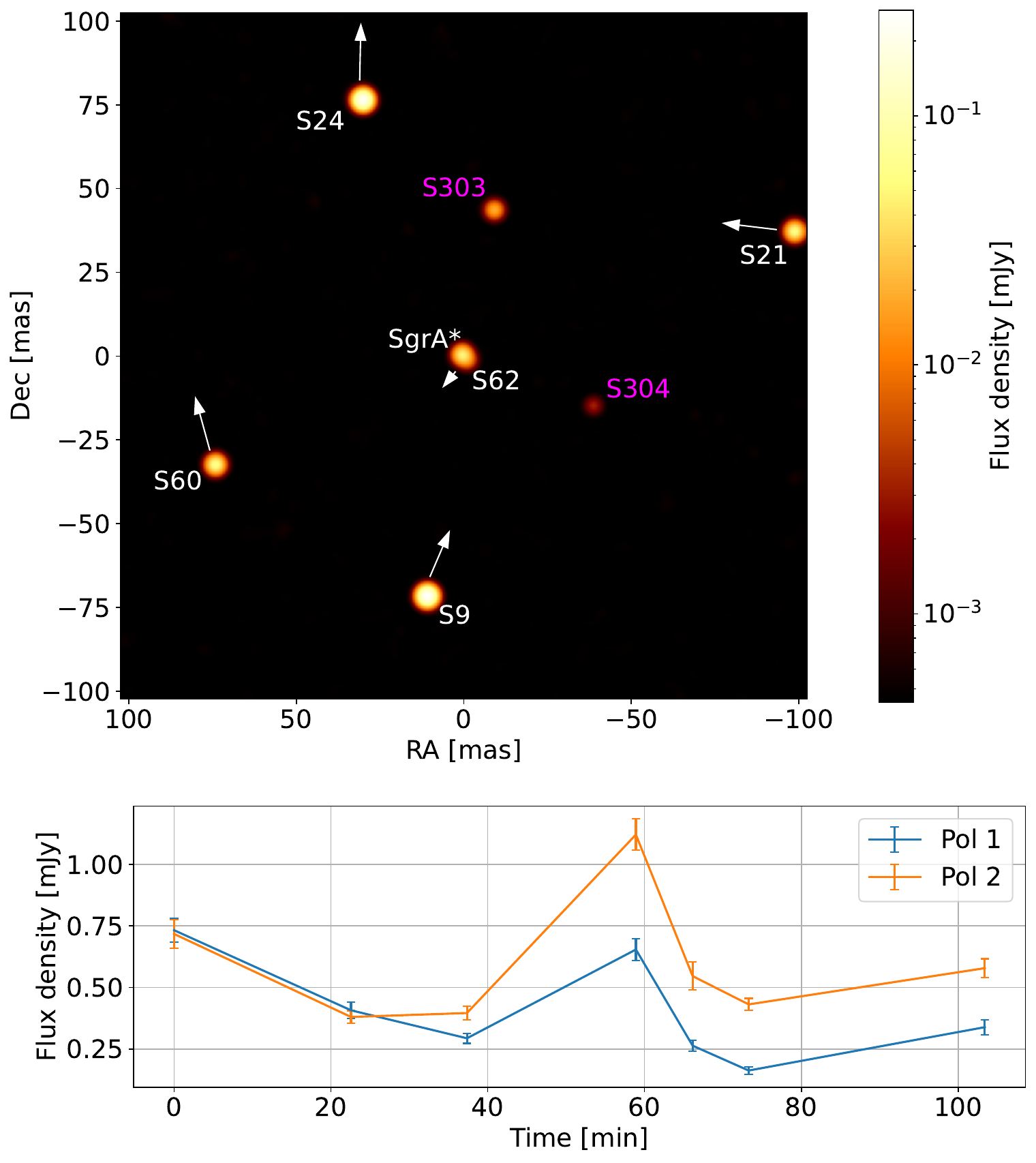}
    \caption{\textit{Upper panel:} GRAVITY+ image of the Galactic Center reconstructed with $\text{G}^{\text{R}}$ from visibility phases and amplitudes. White-labeled stars are modeled explicitly, while the magenta-labeled stars are inferred in a non-parameterized model. The sources S303 and S304 are faint, not yet discovered stars, which were consistently detected across multiple observing seasons (cf. Fig. \ref{fig_Mosaic}). White arrows indicate the relative proper motion of the stars in the field of view.\\
    \textit{Lower panel:} Corresponding light curve of SgrA* for the two linear polarizations.}
    \label{fig_GPAO_image}
\end{figure}

The metrology system in GRAVITY \citep{lippa_metrology_2016} links in dual-field mode the science to the fringe tracker interferogram. This enables minute-long integrations on one hand and allows the measurement of the visibility phase. Contrary to closure phases, visibility phases are position-dependent, hence provide astrometry with respect to the fringe tracker object and offer a reduced noise level.\\
We extend the capabilities of $\text{G}^{\text{R}}$ from reconstructing an image from closure phases and visibility amplitudes \citep{gravity_collaboration_deep_2022} to reconstruct visibility phases and amplitudes, effectively making use of the complex visibilities.\\
We demonstrate this new reconstruction mode by applying $\text{G}^{\text{R}}$ to recent GRAVITY Galactic Center data of March 2025 (subsection \ref{subsection_gpao_data}). In this particular epoch, we expect the stars S9, S24, S60, S21, and S62 to be in the FOV, besides SgrA*. We impose a position prior on all stars and SgrA*, which reflects current knowledge on their positions (Appendix \ref{appendix_gc_model}).\\
We apply an inference scheme (Appendix \ref{subsec_inference_strategy}) that is fitting SgrA*, all prespecified stars, and a background component in the first three global iterations, before the faint object model is inferred. The image is converging first in a low-resolution grid with a pixel size of $\SI{1.6}{mas}$, and is transitioning after 11 global iterations to a high-resolution grid with a pixel size of $\SI{0.8}{mas}$, while keeping the FOV constant. This approach follows the refinement model as outlined in \cite{edenhofer_sparse_2022}. The overall convergence takes in total \SI{15}{min}, and the evolution of the image over the course of a reconstruction is depicted in Fig. \ref{fig_GPAO_image_inference}.\\
The image reconstruction of this dataset with the initial version of $\text{G}^{\text{R}}$ \citep{gravity_collaboration_deep_2022} takes 9.5 times as long using the same computational setup. Before inferring the actual image, error scalings are adapted manually beforehand, which demands even more applications of the code, amounting to at least about 47.5 times the runtime of the revised version of $\text{G}^{\text{R}}$.\\
A total of 10 reconstructions with varying initial random seeds are carried out, and we select the best converged result. The corresponding image with the inferred lightcurve of SgrA* is shown in Fig. \ref{fig_GPAO_image}.
The light curve of SgrA* is overall fairly dim with a mean flux density of $\SI{1.00\pm 0.02}{\milli\text{Jy}}$ corresponding to a K-band magnitude of $m_{\text{K}}=\SI{17.11\pm 0.02}{}$, and moderate variability. The flux density of SgrA* in the reconstructed image is the mean across exposures (time) and the sum over polarizations.\\
Besides bright stars, two faint, yet undiscovered sources are detected to the North and to the West of SgrA*. We do not find any known stars that would coincide with their positions in March 2025 and subsequently label them as S303 and S304, respectively.\\
We find the background noise of the image to be low, with a flux density RMS of $\SI{0.44}{\nano\text{Jy}}$, corresponding to a K-band magnitude of $m_{\text{K}}=\SI{25.5}{}$.\\
This image highlights the ability of $\text{G}^{\text{R}}$ to find and reconstruct multiple, not explicitly modeled stars (in this case S303 and S304), and to infer a lightcurve of SgrA* simultaneously while being provided a dataset of typical size.
\subsection{First interferometric mosaic image in the near-infrared}
\begin{figure*}[hbt!]
    \centering
    \includegraphics[width=\linewidth]{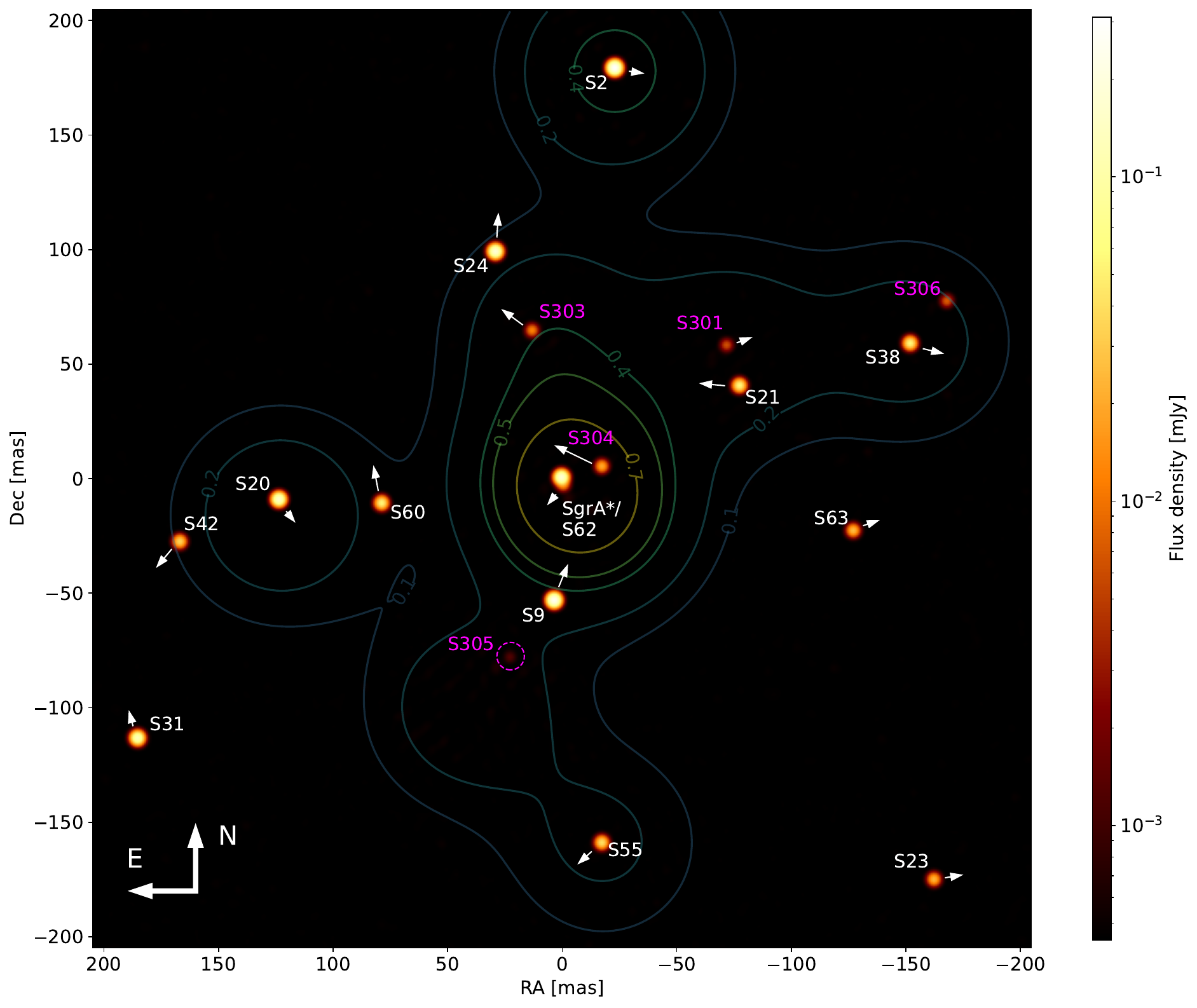}
    \caption{First interferometric mosaic in the NIR, reconstructed with $\text{G}^{\text{R}}$ from 88 GRAVITY+ exposures of the Galactic Center, acquired in August 2025 (Tab. \ref{tab_mosaic_dataset}). In this mosaic, nine spatially separated datasets have been reconstructed jointly into a single image at a scale of $\SI{409.6}{mas}\times\SI{409.6}{mas}$, or $\simeq\SI{3390}{AU}\times\SI{3390}{AU}$ with a resolution of $\SI{0.8}{mas}$ per pixel. We explicitly modeled all white-labeled stars with a parameterized model, while the magenta-labeled stars have been inferred by a non-parameterized model. These sources are faint, yet undiscovered stars in the Galactic Center (except for S301), of which S303 and S304 could be found consistently across multiple observing seasons. S305 and S306 are currently being followed up on. The white arrows indicate the relative proper motions of the stars in the field of view. The contours depict the theoretical relative sensitivity within the image representing the throughput profile of GRAVITY's single-mode fibers weighted with the square root of the number of used exposures for the respective pointing.}
    \label{fig_Mosaic}
\end{figure*}

The ability of GRAVITY to measure visibility phases not only facilitates the reconstruction of deeper images, but yields an observable which is sensitive to the absolute position on sky (with respect to the fringe tracker star GC IRS 16C \citep{abuter_first_2017}). Regarding image reconstruction, it becomes possible to reference multiple, spatially separate acquired pointings/datasets to each other and to reconstruct them jointly into a single image. Contrary to reconstructing individual datasets and assembling the corresponding images into a larger image \citep{gravity_collaboration_deep_2022}, a true mosaic, as practiced in radio interferometry, reconstructs the information in overlapping pointings as well. The added information in these regions leads to suppression of noise and ultimately to additional depth. Mosaicking effectively increases the FOV of the interferometer while profiting from GRAVITY's supreme image resolution.\\
We apply the mosaic forward model (subsection \ref{subsection_mosaicing}), which is shown in detail for the special case of three pointings in Fig. \ref{fig_Mosaic_model}. In theory, this scheme can be extended to an arbitrary number of pointings, and the only limiting factor is computational expense. Given a fixed number of exposures per pointing and a fixed extent of the FOV, the computational effort scales approximately linearly with the number of pointings.\\
In practice, we demonstrate the power of this approach by jointly reconstructing in total nine pointings scattered across a FOV of $\simeq\SI{400}{mas}\times\SI{400}{mas}$. These nine pointings contain 88 GRAVITY exposures acquired across three nights; an overview of the data used is given in Tab. \ref{tab_mosaic_dataset}. We expect a total of 13 bright stars within the FOV and model them explicitly (Appendix \ref{appendix_gc_model}). We apply $\text{G}^{\text{R}}$ with the inference scheme outlined in Appendix \ref{subsec_inference_strategy}. The convergence of the method with the mosaic model in place takes in total $\simeq\SI{31}{\hour}$, i.e., $\sim124$ times longer compared to a single pointing reconstruction. We reconstruct the mosaic a total of seven times with varying initial random seeds and select the cleanest result. The corresponding image, the first true interferometric mosaic in the near-infrared, is presented in Fig. \ref{fig_Mosaic}. Besides the expected bright sources, we recover the faint stars S303 and S304, which were detected in the March 2025 dataset (subsection \ref{subsection_rec_visphi}). Moreover, two faint sources to the far North-West of SgrA* are inferred, S301 $\left(m_{\text{K}}\simeq\SI{19.3}{}\right)$, which has been detected across multiple observing seasons \citep{abd_el_dayem_discovery_2026}, and an unknown source S306 $\left(m_{\text{K}}\simeq\SI{19.12}{}\right)$. Additionally, a very faint source was inferred robustly to the South of SgrA* $\left(m_{\text{K}}>\SI{20.0}{}\right)$, which we label as S305. The sources S305 and S306 have not yet come to our attention before the reconstruction of this mosaic, and are currently being followed up on. The mosaic image features a noise level of $\SI{0.44}{\nano\text{Jy}}$ RMS, corresponding to a K-band magnitude of $m_{\text{K}}=\SI{25.5}{}$, i.e., a similar noise floor to the image reconstructed from a single dataset (subsection \ref{subsection_rec_visphi}).
\begin{figure}[t]
    \centering
    \includegraphics[width=\linewidth]{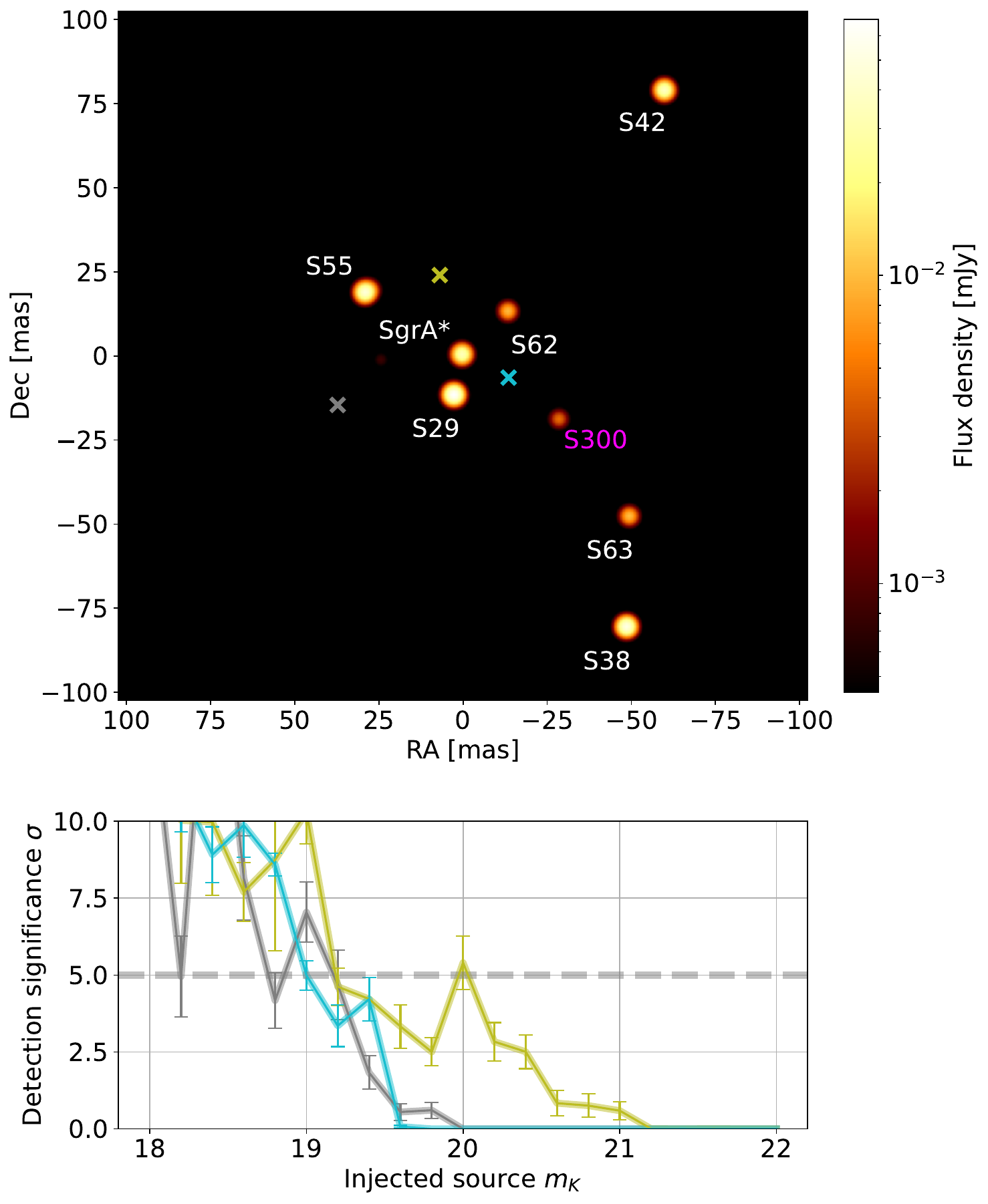}
    \caption{Performance test of $\text{G}^{\text{R}}$ on conventional GRAVITY data.\\\textit{Upper panel:} Mean image from 10 reconstructions of the GRAVITY dataset for 29 May 2021. The faint star S300, first discovered in 2022, is routinely inferred in nine out of 10 reconstructions with high significance and in one reconstruction with low significance.\\
    \textit{Lower panel:} Detection significance of artificially injected sources for three different positions (turquoise, yellow, grey crosses) injected in the object model of the upper image, plotted against the K-band magnitude of the respective injected source. $\text{G}^{\text{R}}$ is able to recover sources up to $m_{\text{K}}=\SI{20.0}{}$. See subsection \ref{subsection_performance_test} for details. Statistical errors on the detection significance are derived from jackknife resampling.}
    \label{fig_injection_map}
\end{figure}
\subsection{Performance test}\label{subsection_performance_test}
\subsubsection{Application to conventional GRAVITY data}
We identify three performance criteria relevant for routine GC observations and the science objective to discover faint stars: robustness, computational speed, and the limiting magnitude in images. To test the performance of $\text{G}^{\text{R}}$, we use a conventional GRAVITY dataset (subsection \ref{subsection_performance_dataset}), which was acquired prior to any upgrades of GRAVITY+.\\
We first quantify how robust the method is inferring an image. Besides SgrA*, we expect six stars in the FOV, which are S55, S29, S62, S42, S63, and S38, and model them explicitly. Moreover, this dataset features the faint star S300, discovered in \cite{gravity_collaboration_deep_2022}, which we do not model explicitly and leave for the image to be inferred. We run $\text{G}^{\text{R}}$ in total 10 times with varying initial random seeds to increase the number of samples and hence improve statistics on the image.\\
The mean over these 10 reconstructions is presented in the upper panel of Fig. \ref{fig_injection_map}. The image features a significant detection of S300, inferring the star in nine of 10 reconstructions. In one reconstruction, $\text{G}^{\text{R}}$ is still inferring S300 but with very low flux density, amounting to a low significance. Besides S300, there is only one other source present in the image, distinct from noise, located to the East of SgrA*. It is inferred in two out of 10 reconstructions, overall decreasing its detection significance to $1.1\sigma$ (Appendix \ref{subsection_statistical_significance}). With its K-band magnitude of $m_{\text{K}}\simeq\SI{20.1}{}$, it is on the detection limit of conventional GRAVITY data in the GC, and we follow up on it with current observations. The background noise level, that is, the RMS over the entire image with SgrA* and all known stars excluded, amounts to $m_{\text{K}}=\SI{25.4}{}$ or $\SI{0.49}{\nano\text{Jy}}$.\\
We use the model of the reconstructed image (upper panel in Fig. \ref{fig_injection_map}) as the basis to conduct injection tests with artificial sources and to ultimately arrive at a quantified detection limit for GRAVITY.\\
\subsubsection{Limiting K-band magnitude of GRAVITY in the Galactic Center}\label{subsubsection_limiting_K_band_magnitude}
To determine the limiting magnitude of GRAVITY images in the GC, we inject artificial point sources into the same conventional GRAVITY dataset (subsection \ref{subsection_performance_dataset}) and try to recover them. The sources are injected by first computing the corresponding instrument response of GRAVITY to the reconstructed image (upper panel in Fig. \ref{fig_injection_map}) and subsequently subtracting the data. This should ideally yield pure noise, and we add back the response model augmented by an artificial source at a specific position and flux density. We gradually decrease the injected flux density of the artificial source until the source can no longer be detected.\\
We inject sources at three different positions with projected distances to SgrA* of $\SI{15}{mas}$, $\SI{25}{mas}$, and $\SI{40}{mas}$. Along these annuli, we choose dark patches with maximum projected distance to nearest-neighbor stars (see yellow, magenta, and cyan crosses in Fig. \ref{fig_injection_map}, respectively). We progressively increase their K-band magnitude from $m_{\text{K}}=18.0$ to $m_{\text{K}}=22.0$ in steps of 0.2 and reconstruct only one injected source at a time, while we expect S300 to be inferred along.\\
The detection significance for the recovered injected sources is presented in the lower panel of Fig. \ref{fig_injection_map}. We apply the definition of the statistical significance given in Appendix \ref{subsection_statistical_significance} and find that injected sources can be detected up to $m_{\text{K}}=\SI{20.0}{}$ with high significance. In \cite{gravity_collaboration_deep_2022}, a limiting magnitude of $m_{\text{K}}=\SI{19.7}{}$ was reported for a dataset containing six more exposures than the performance dataset (subsection \ref{subsection_performance_dataset}). Assuming that the limiting flux density is proportional to the square root of reconstructed datapoints, this constitutes a gain in depth by a factor of 1.65. Correspondingly, the limiting magnitude of $\text{G}^{\text{R}}$ increased by $\Delta m_{\text{K}}=\SI{0.55}{}$.
\section{Discussion}\label{section_discussion}
With the presented developments in place, $\text{G}^{\text{R}}$ is now faster, more robust, recovers deeper images, reconstructs complex visibilities, and facilitates mosaicking in the near-infrared. There are, nevertheless, a number of limitations to this method, regarding inferred error estimates on physical parameters, noise statistics of GRAVITY, and overall restrictions by the implemented model of the Galactic Center.
\subsection{Statistical error estimates}
The underlying inference method in $\text{G}^{\text{R}}$, MGVI \citep{knollmuller_metric_2020}, makes a fundamental assumption that has an impact on error quantification; it presumes that the covariance of the approximating posterior, the multivariate normal distribution, is sufficiently described by the inverse Fisher information matrix \citep{knollmuller_metric_2020}. Hence, the standardization function (Appendix \ref{subsection_standardization}), which maps latent variables to physical parameters describing the object model, is supposed to be adequately linear \citep{knollmuller_metric_2020}. This allows MGVI to scale in computational cost only linearly with the number of parameters \citep{knollmuller_metric_2020}.\\
The full expression for the inverse covariance of the approximating posterior is given as \citep{knollmuller_metric_2020},
\begin{flalign}
    & \Theta^{-1}\ =\ \left<\frac{\partial\mathcal{H}\left(d,\xi\right)}{\partial\xi}\frac{\partial\mathcal{H}\left(d,\xi\right)}{\partial\xi^{\dagger}}\right>_{\mathcal{N}\left(\xi|\bar{\xi},\Theta\right)} - \\
    & \left<\frac{1}{\mathcal{P}\left(d,\xi\right)}\frac{\partial^{2}\mathcal{P}\left(d,\xi\right)}{\partial\xi\partial\xi^{\dagger}}\right>_{\mathcal{N}\left(\xi|\bar{\xi},\Theta\right)}.
\end{flalign}
By definition, the covariance must be positive definite, which is not the case in the presence of the second summand. For this reason, the second term is neglected in MGVI \citep{knollmuller_metric_2020}, leading to the expression provided in Eq. \ref{eq_approx_cov}. This can result in an underestimation of the covariance, and hence, in the underestimation of errors on physical parameters.\\
In the case of a standardization mapping that is deviating from a linear relationship, the Cram\'er-Rao bound is no longer described solely by the inverse Fisher information matrix and more summands are accumulated, which are ignored in MGVI \citep{knollmuller_metric_2020,barrau_note_2013}. In the case of the faint object model (Appendix \ref{subsection_faint_stars}), where the standardization function is largely non-linear, this, too, can lead to an underestimation of errors.
\subsection{Correlations/noise statistics of GRAVITY data}\label{subsection_correlations}
The model in $\text{G}^{\text{R}}$ is based on assumptions with respect to the noise statistics and correlations in GRAVITY data. Several processes contribute to the noise in GRAVITY data, resulting in the approximation of noise statistics to be adequately normal distributed.\\
Moreover, the likelihood, as implemented in $\text{G}^{\text{R}}$, is presumed to have negligible off-diagonal components in its covariance matrix. This is equivalent to every GRAVITY datapoint being entirely uncorrelated, which is not the case for GRAVITY \citep{kammerer_increasing_2020}. By taking the full covariance matrix into account, \cite{kammerer_increasing_2020} reports that, in their application of finding companions, the detection limit can be raised by a factor $\sim\SI{3.5}{}$. Applying this value to the detection limit from our injection tests, taking the full covariance matrix in $\text{G}^{\text{R}}$ into account, could improve the limiting magnitude for GRAVITY in the GC from $m_{\text{K}}=\SI{20.0}{}$ up to $m_{\text{K}}\simeq\SI{21.4}{}$.
\subsection{Model limitations}
$\text{G}^{\text{R}}$ is designed to only reconstruct point sources, as GRAVITY does not resolve the finite angular extent of stars in the Galactic Center. This model is limiting in two ways; in the event of SgrA* flaring, GRAVITY is able to resolve motion in the accretion zone \citep{abuter_polarimetry_2023}, which makes a stationary, point source model of SgrA* imprecise. Secondly, any large, cohesive, faint flux distribution is omitted so far. Stellar counts in the central $\SI{9}{as}$ in the GC \citep{gallego-cano_photometric_2024} suggest a cusp, that is, a diverging power law towards SgrA*. On scales of $\simeq\SI{100}{mas}\times\SI{100}{mas}$, accessible to GRAVITY, an extrapolation to the very center is unphysical. If the unresolved flux contribution at the inner scales is radially flat enough, it does not contribute to complex visibilities other than an overall lower level of visibility amplitudes. Nevertheless, for future work, a background object model, parametrized by a radially symmetric slope, could be added to the overall object model.\\
The slope of the power spectral density (PSD) of SgrA*'s lightcurves and the break frequency are varying \citep{yusef-zadeh_nonstop_2025}. It is left for future research to explore a model that adds slope and break frequency to the set of inferred parameters.\\
Large-scale mosaics, as demonstrated in Fig. \ref{fig_Mosaic}, require multiple nights for data acquisition. For stars close to peri-center, or generally speaking, for stars with considerably high proper motion, a stationary object model could not suffice. It would then be appropriate to extend the model for every point source (beyond SgrA*) in the time domain.\\
Finally, $\text{G}^{\text{R}}$ could be adapted to measure intrinsic flux densities of stars in the GC instead of referencing the entire image to the brightest star in the FOV. This would remove the dependence on prior flux density measurements, acquired with a single $\SI{8}{\text{m}}$-class telescope.
\section{Conclusion}\label{section_conclusion}
We present in this work a revision of the image reconstruction method GRAVITY-RESOLVE $\text{G}^{\text{R}}$ \citep{gravity_collaboration_deep_2022}, which improves speed, robustness, provides the applicability to other data products, and the possibility to perform mosaicking as practiced in radio interferometry \citep{sault_approach_1996}. This method acts as a basis for future Galactic Center observations with GRAVITY+, offering a \textit{live view} of the GC and the possibility for both image reconstruction and model fitting of complex visibilities in near-infrared interferometry.
\subsection{Fast and reliable image reconstruction method for GRAVITY+}
GRAVITY+ is the instrument upgrade to GRAVITY, comprising the beam combiner instrument and the VLTI infrastructure \citep{collaboration_gravity_2022}. With the implementation of extreme (advanced) adaptive optics in the fall of 2024 and with equipping each 8m UT telescope of the VLTI with a laser guide star (LGS), the GRAVITY+ upgrade comes to full fruition. This yields a higher Strehl ratio effectively, and hence a higher throughput of light to the beam combiner instrument. With respect to the Galactic Center science case, a higher throughput means additional depth in images. Moreover, it becomes possible to observe in medium spectral resolution mode. In this mode, the number of available datapoints to reconstruct is effectively increased, gaining additional depth and effectively extending the FOV of images \citep{eisenhauer_advances_2023}. This comes, however, with a surplus of computational cost as the number of spectral channels increases from 14 to 233. This is where a shortcut within $\text{G}^{\text{R}}$ (Appendix \ref{subsection_reduced_number_of_FFT}) comes into place to circumvent the computation of complex visibilities within the forward model for every spectral channel and instead scale visibilities to account for systematics. With the LGS fully operational since late 2025, this method comes in time as a reliable tool for imaging the Galactic Center with GRAVITY+. With the LGS in place, we are able to discover fainter stars in the Galactic Center and follow up on them longer with an increased FOV in medium spectral resolution.
\subsection{Future prospects}
If one would accept substantial changes to the underlying model, the optimization method, or the used programming language altogether, there is still ample room for improvement with respect to speed and convergence of the method.\\
A potential prospect for a more efficient model would be to apply the concept of compressed sensing \citep{baron_inference_2020}, which eliminates redundant elements of a model, both saving degrees of freedom as well as computational cost. For the case of $\text{G}^{\text{R}}$, this would mean to abandon the concept of an uncorrelated field of pixels, and instead model faint stars explicitly as point sources and let them find their respective position and K-band magnitude within the FOV. Moreover, since an average GRAVITY dataset is comprised of $\mathcal{O}\left(1000\right)$ data points, the inference problem would be overdetermined, making the application of the optimization method MGVI \citep{knollmuller_metric_2020} highly efficient. This would be, however, no longer imaging, but model fitting, and comes with the risk of potentially locking one or more of these stars in a side-lobe of the dirty image, resulting in wrong astrometry.\\
The optimization method MGVI \citep{knollmuller_metric_2020} makes the strong assumption that every parameter of the inference problem is approximately normally distributed, or that the inverse transform sampling relationship is largely linear. This ensures the linear scaling of computational expense with the number of parameters, but is a limiting factor when it comes to highly skewed distributions like the one for the faint object model (Appendix \ref{subsection_faint_stars}). We propose to investigate other optimization methods like Hamiltonian Monte Carlo \citep{duane_hybrid_1987}, which is available in the NIFTy package \citep{lienhard_hmcf_2018} and is able to retrieve the true posterior \citep{knollmuller_metric_2020}.\\
With respect to the limiting magnitude in GRAVITY images, significant progress could be achieved by taking the full covariance matrix, that is, correlations between GRAVITY datapoints, into account (subsection \ref{subsection_correlations}). Applying the increase in detection limit reported in \cite{kammerer_increasing_2020} to GC data, the limiting magnitude could be improved to $m_{\text{K}}\simeq\SI{21.4}{}$ just in the case of conventional GRAVITY data, prior to GRAVITY+.\\
Finally, the modular structure of $\text{G}^{\text{R}}$ allows for the exchange or extension of any of its model constituents, allowing, for example, to link pixels in the faint object model via a known spatial correlation to reconstruct extended objects.
\begin{acknowledgements}
This work was supported by Paris Île-de-France Region; the French National Research Agency (ANR) under grant
ANR-23-EDIR-0003 (GRAFITY); the "Action Thématique Gravitation Références Astronomie Métrologie"
(ATGRAM), the "Action Thématique Phénomènes Extrêmes et Multimessagers"
(ATPEM), and the "Action Thématique Cosmologie et Galaxies" (ATCG), of
CNRS/INSU, with co-funding by CNRS/IN2P3, CNRS/INP, CEA and CNES. JS acknowledges funding from the Deutsche Forschungsgemeinschaft (DFG, German Research Foundation) under its Emmy-Noether Programm (STA 1955/1-1, Projektnummer 545534254).
\end{acknowledgements}
\bibpunct{(}{)}{;}{a}{}{,} 
\bibliographystyle{aa}
\bibliography{references}

\begin{appendix}
\section{Model of the Galactic Center}\label{appendix_gc_model}
The object model of the Galactic Center in $\text{G}^{\text{R}}$ features the spatial, photometric, spectral, and temporal dimensions. It consists of dedicated models for SgrA*, known bright sources within the field of view (FOV), and a model for faint, yet undiscovered sources.\\
SgrA* and bright sources are explicitly implemented as a parameterized model, while the faint object model is represented by a non-parameterized model, i.e., a continuous field of pixels. SgrA* is the sole model component to exhibit time variability, accounting for its strong fluctuations in flux density, during quiescent and flaring episodes \citep{do_near-infrared_2009,dodds-eden_two_2011,witzel_source-intrinsic_2012,witzel_variability_2018,chen_consistency_2019,abuter_flux_2020,witzel_rapid_2021,abuter_constraining_2021,weldon_near-infrared_2023,yusef-zadeh_nonstop_2025}.\\
In the following, we present the three components based on astrophysical arguments.
\subsection{SgrA*}
\subsubsection{Position}
We define the position of SgrA* with respect to the phase center, i.e., the position where the visibility phase is zero. While the intrinsic motion of SgrA* on timescales of monthly observing campaigns is negligible \citep{plewa_pinpointing_2015}, its position is subject to instrumental systematics. Potential small pointing offsets of fibers in the GRAVITY beam combiner, and any residual drifts of the science object with respect to the fringe tracker star, are accounted for by modeling the position of SgrA* $\boldsymbol{s}_{\text{SgrA*}}$ with a normal distribution in which the expected position is reflected by the mean $\boldsymbol{\mu}_{\text{SgrA*}}$ and any deviation by a standard deviation $\sigma_{\text{SgrA*}}\simeq\SI{0.5}{mas}$ in both RA and Dec:
\begin{equation}
    \boldsymbol{s}_{\text{SgrA*}}\ \sim \ \mathcal{N}\left(\boldsymbol{\mu}_{\text{SgrA*}},\sigma_{\text{SgrA*}}^2\mathbf{1}\right)
\end{equation}
\subsubsection{Flux density}
The flux density of SgrA* follows a log-normal distribution with an additional tail towards higher fluxes to account for its flaring episodes \citep[e.g.][]{abuter_flux_2020}. We approximate this distribution by modeling the flux density as an inverse gamma distribution following \cite{abuter_flux_2020}. Additionally, we invoke the temporal correlation within the lightcurve of SgrA*. In the forward model of $\text{G}^{\text{R}}$, we apply the following steps to arrive at a temporally correlated lightcurve model of SgrA* like in \cite{timmer_generating_1995,witzel_source-intrinsic_2012,witzel_variability_2018,arras_comparison_2021,arras_variable_2022,edenhofer_sparse_2022}, exploiting the Wiener-Khintchin theorem \citep{wiener_extrapolation_1949,khintchine_korrelationstheorie_1934}:\\ \\
1. Draw samples $\xi^{*}$ from a complex standard normal distribution for $N$ temporal frequencies $f_{\text{n}}$, with $n=\left\{1, ..., N\right\}$, for two linear polarizations separately,
\begin{equation}
    \xi^{*}\left(f_{\text{n}}\right)\ \sim\ \mathcal{N}_{\mathcal{C}}\left(0, \mathbf{1}\right).
\end{equation}
2. Weight samples by the square root of the power spectral density (PSD) $P$ at the corresponding frequency $f_{\text{n}}$,
\begin{equation}
    \xi^{*}\left(f_{\text{n}}\right)\sqrt{P\left(f_{\text{n}}\right)}.
\end{equation}
3. Apply the inverse of the one-dimensional, discrete, real domain Fourier Transform,
\begin{equation}\label{eq_temporal_samples_non_normalized}
    \xi^{*}\left(t_{\text{m}}\right)\ =\ \mathcal{F}^{-1}\left[\xi^{*}\left(f_{\text{n}}\right)\sqrt{P\left(f_{\text{n}}\right)}\right],
\end{equation}
to obtain $M$ temporal samples $\xi^{*}\left(t_{\text{m}}\right)$, with $m=\left\{1, ..., M\right\}$. Since the frequency samples in Step 1 are drawn independently, the corresponding covariance matrix is diagonal. It can be shown that the covariance matrix of temporal samples then evaluates to $S_{t_{\text{i}},t_{\text{j}}}\ =\ \mathcal{F}^{-1}\left[P\left(f_{\text{n}}\right)\right]\left(t_{\text{i}}-t_{\text{j}}\right)$, assigning a variance of $\sigma_{t_{\text{m}}}^2\ =\ \mathcal{F}^{-1}\left[P\left(f_{\text{n}}\right)\right]\left(0\right)$ to the distributions of temporal samples.\\
\begin{figure}
    \centering
    \includegraphics[width=\linewidth]{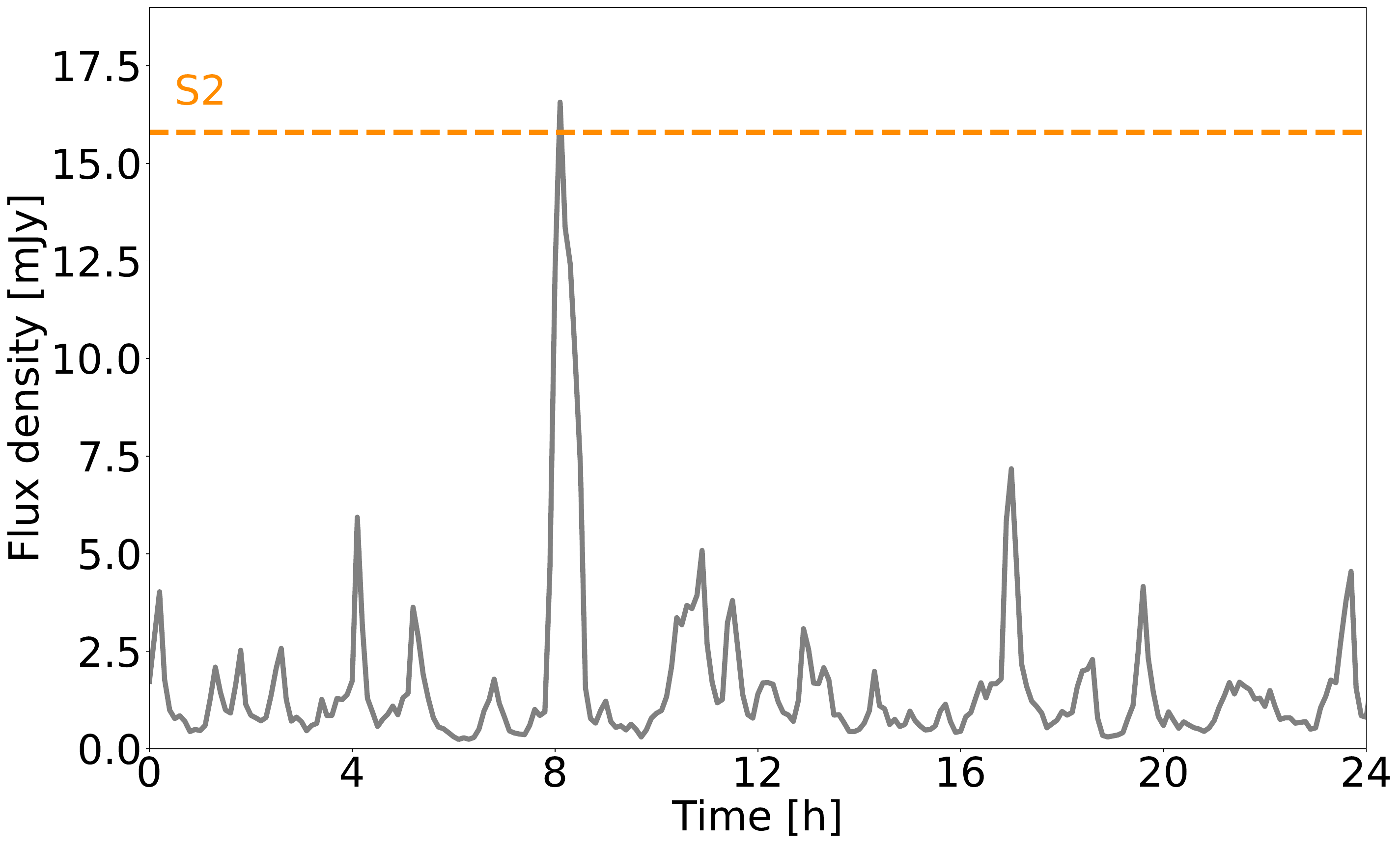}
    \caption{Random instance of the lightcurve model of SgrA* as implemented in $\text{G}^{\text{R}}$ over the course of $\SI{24}{h}$. The orange dashed line represents the dereddened flux density of the star S2 \citep{abuter_flux_2020}. This model incorporates both the measured power spectral density and the flux density distribution in the lightcurve of SgrA*.}
    \label{fig_SgrA_flux_time}
\end{figure}
4. Divide Eq. \ref{eq_temporal_samples_non_normalized} by $\sigma_{t_{\text{m}}}$ to obtain samples in the time domain which follow a standard normal distribution,
\begin{equation}
    \xi^{*}\left(t_{\text{m}}\right)\ =\ \frac{\mathcal{F}^{-1}\left[\xi^{*}\left(f_{\text{n}}\right)\sqrt{P\left(f_{\text{n}}\right)}\right]\cdot\sqrt{M}}{\sqrt{2\cdot\mathcal{F}^{-1}\left[P\left(f_{\text{n}}\right)\right]\left(0\right)}}.
\end{equation}
The weight $\sqrt{\frac{1}{2}}$ accounts for the fact that the real and imaginary parts of the frequency samples $\xi^{*}\left(f_{\text{n}}\right)$ in Step 1 are drawn independently from two standard normal distributions. We ensure that the normalization of the inverse Fourier Transform has no impact on the distribution of temporal samples by applying $\sqrt{M}$ to the numerator.\\
5. Finally, choose those temporal samples $\xi^{*}\left(t_{\text{m}}\right)$ which are closest to the time stamps of exposures during an observation. These samples are subsequently transformed to samples of an inverse gamma distribution \citep{abuter_flux_2020},
\begin{equation}
    F_{\text{SgrA*}}\left(t_{\text{m}}\right)\ \sim \ \text{Inv-Gamma}\left(\alpha = 1.76, \beta = 1.49\right),
\end{equation}
approximating the flux density distribution of SgrA* \citep{abuter_flux_2020}.\\
A random realization of this forward model is plotted in Fig. \ref{fig_SgrA_flux_time}, and its corresponding power spectrum is depicted in Fig. \ref{fig_SgrA_psd_frequency}, which recovers the imposed PSD.
\begin{figure}
    \centering
    \includegraphics[width=\linewidth]{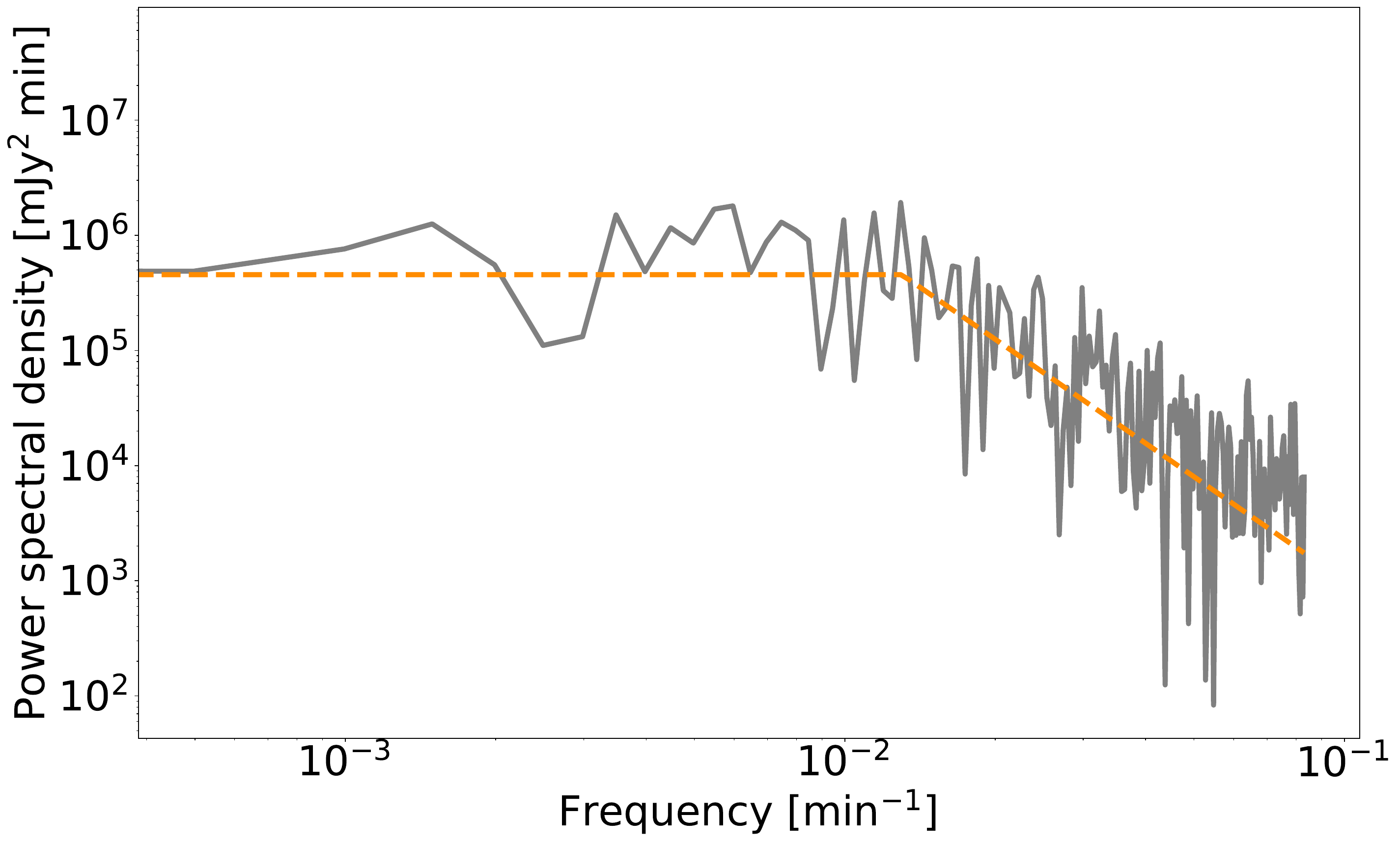}
    \caption{Power spectrum of the exemplary lightcurve model of SgrA* in Fig. \ref{fig_SgrA_flux_time}. The orange dashed line represents the invoked power spectrum from \citep{yusef-zadeh_nonstop_2025}.}
    \label{fig_SgrA_psd_frequency}
\end{figure}
\subsubsection{Spectral energy distribution}
We model the spectral energy distribution ($F_{\lambda}=F_{\text{SgrA*}}\cdot f_{\alpha_{\text{SgrA*}}}\left(\lambda\right)$) of SgrA* according to \cite{gravity_collaboration_deep_2022}, following a power law with a slope of $-\alpha_{\text{SgrA*}}-1$,
\begin{equation}\label{eq_sgra_spec_dist}
    f_{\alpha_{\text{SgrA*}}}\left(\lambda\right)\ =\ \left(\frac{\lambda}{\SI{2.2}{\mu\meter}}\right)^{-\alpha_{\text{SgrA*}}-1}.
\end{equation}
The spectral index $\alpha_{\text{SgrA*}}$ is variable with time and dependent on the flux density of SgrA* \citep{yusef-zadeh_nonstop_2025}. We account for the variability by imposing a normal distribution on $\alpha_{\text{SgrA*}}$,
\begin{equation}
    \alpha_{\text{SgrA*}}\ \sim\ \mathcal{N}\left(\mu_{\alpha_{\text{SgrA*}}}, \sigma_{\alpha_{\text{SgrA*}}}^2\right),
\end{equation}
with a sufficiently large standard deviation of $\sigma_{\alpha_{\text{SgrA*}}}=\SI{3}{}$ around a mean of $\mu_{\alpha_{\text{SgrA*}}}=\SI{0}{}$, as SgrA* exhibits a shallow spectrum in the frequency domain \citep{yusef-zadeh_nonstop_2025}.
\subsection{Bright stars}
In $\text{G}^{\text{R}}$, we explicitly model every star with a K-band magnitude $m_{\text{K, bright}}\ <\ 18.9$ \citep{abuter_detection_2021} within the GRAVITY FOV, i.e., stars only accessible to single $\SI{8}{\meter}$-class telescopes. This boundary is set by the star S62, which is the faintest star pursued with the NACO instrument \citep{abuter_detection_2021}. We fix the flux density of the brightest source as the van Cittert-Zernike theorem (Eq. \ref{eq_van_cittert-zernike}), and correspondingly, the response equation of GRAVITY (Eq. \ref{eq_reponse_aberr}) is scale invariant with respect to flux density and hence facilitates the inference (cf. subsection \ref{subsec_inference_strategy}).
\subsubsection{Position}
As in the object model of SgrA*, we refer positions of known stars in the GC to the phase center. Contrary to SgrA*, their positions are not only subject to small instrumental systematic effects, but also to uncertainties in stellar orbits due to limited time coverage. We also impose a normal distribution on the position of bright stars,
\begin{equation}
    \boldsymbol{s}_{\text{Bright}}\ \sim\ \mathcal{N}\left(\boldsymbol{\mu}_{\boldsymbol{s}_{\text{Bright}}},\sigma_{\boldsymbol{s}_{\text{Bright}}}^2\mathbf{1}\right).
\end{equation}
The mean $\boldsymbol{\mu}_{\boldsymbol{s}_{\text{Bright}}}$ of the position prior is set to a predicted position relative to SgrA* based on relativistic orbit models. We expect for almost all bright stars in the central $\simeq\SI{400}{}\times\SI{400}{mas}$ to be able to predict their positions with $\sigma_{\boldsymbol{s}_{\text{Bright}}}\simeq\SI{0.5}{mas}$ in both RA and Dec. From experience, the depth of a reconstructed image is directly dependent on the inverse width of the position prior, linking a better knowledge of the orbit ultimately to a better image.
\subsubsection{Flux density}
\begin{figure}
    \centering
    \includegraphics[width=\linewidth]{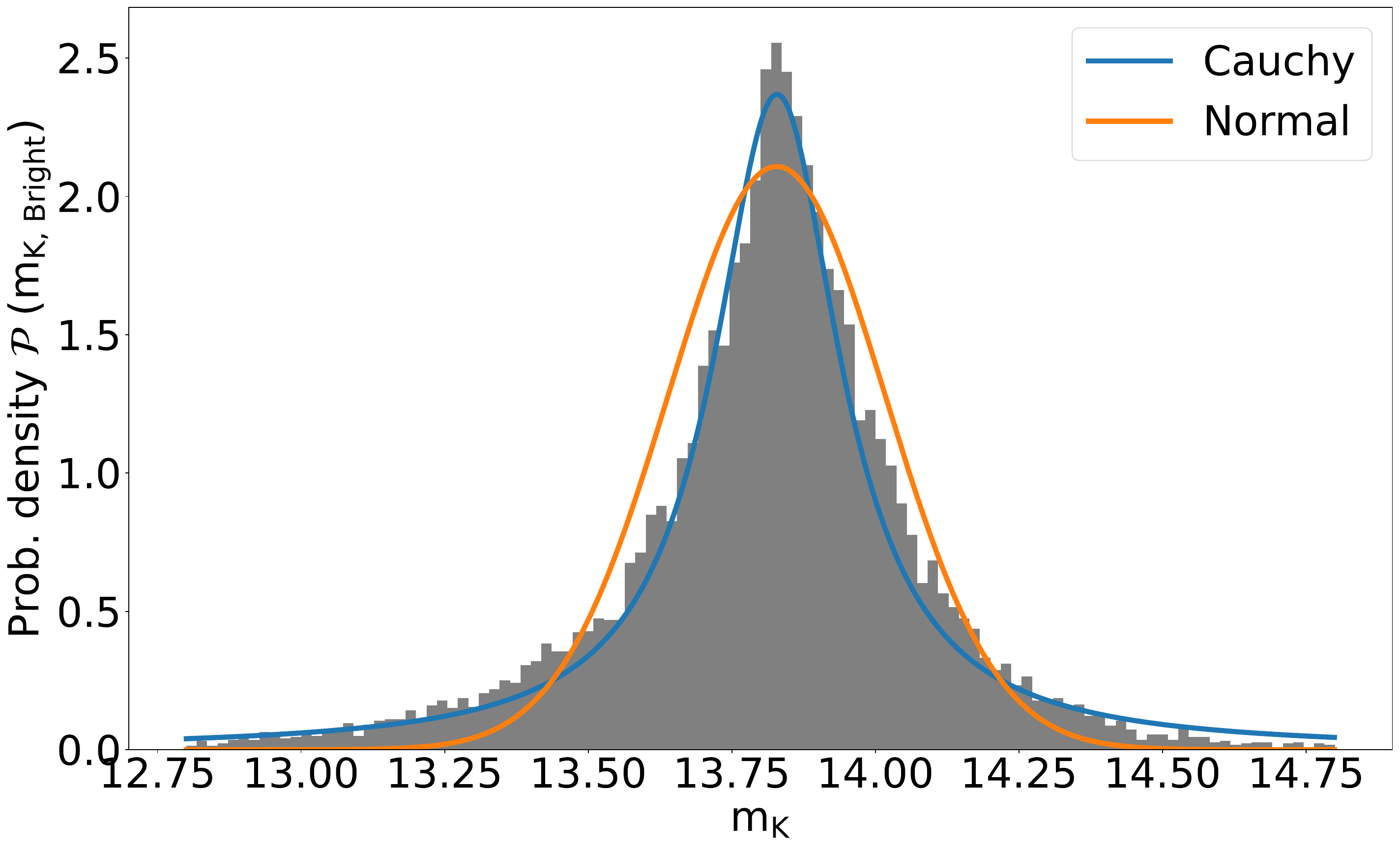}
    \caption{Distribution of K-band magnitude samples of all known stars in the GC \citep{gillessen_update_2017} normalized to the magnitude of the star S2. It can be described with a Cauchy distribution, while the normal distribution serves as a sufficiently adequate approximation.}
    \label{fig_Bright_mK_pdf_paper_new}
\end{figure}
\begin{figure}[H]
    \centering
    \includegraphics[width=\linewidth]{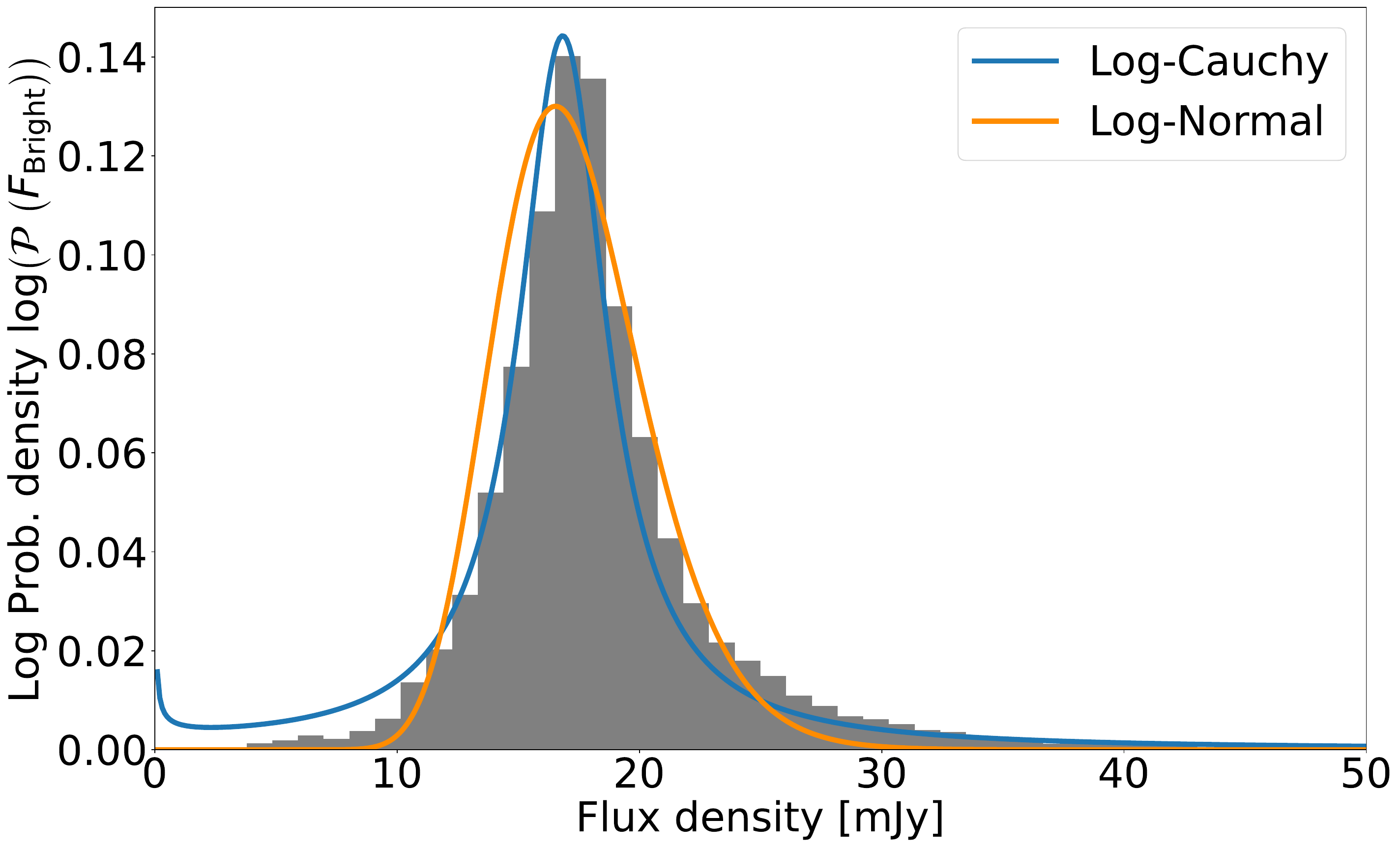}
    \caption{Histogram of the K-band magnitude samples (Fig. \ref{fig_Bright_mK_pdf_paper_new}) transformed to flux densities with its corresponding model (orange line), as applied for bright stars in the GC. The fit of a Log-Cauchy to the histogram yields a poor description for low flux densities and diverges towards zero.}
    \label{fig_Bright_flux_pdf}
\end{figure}

Over the course of the past $\simeq\SI{30}{}$ years, we have monitored the photometry of a couple of dozen stars \citep{gillessen_update_2017}, with both single-telescope observations, as well as interferometry \citep{drescher_probing_2025}. In line with the central limit theorem, we expect the photometric samples to be normally distributed,
\begin{equation}
    m_{\text{K}}\ \sim\ \mathcal{N}\left(m_{\text{K},\text{mean}},m_{\text{K},\text{std}}^2\right),
\end{equation}
with $m_{\text{K},\text{mean}}$ the mean K-band magnitude of the corresponding star and its standard deviation $m_{\text{K},\text{std}}$. A transform to de-reddened flux densities yields a log-normal distribution naturally,
\begin{equation}
    F_{\text{Bright}}\ \sim\ 10^{0.4\left(m_{\text{K},\text{norm}}-\left(\xi\cdot m_{\text{K},\text{std}}+m_{\text{K},\text{mean}}\right)\right)}F_{\text{norm}}\ \overset{!}{=}\ e^{\mu+\sigma\cdot\xi},
\end{equation}
where the parameter $\mu$ can be identified with
\begin{equation}
    \mu\ \coloneqq\ 0.4\ \text{ln}\left(10\right)\left(m_{\text{K},\text{norm}}-m_{\text{K},\text{mean}}\right)+\text{ln}\left(F_{\text{norm}}\right),
\end{equation}
and the parameter $\sigma$ correspondingly with
\begin{equation}
    \sigma\ \coloneqq\ \left|-0.4\ \text{ln}\left(10\right)\ m_{\text{K},\text{std}}\right |.
\end{equation}
An example of this model for the star S2 is illustrated in Fig. \ref{fig_Bright_flux_pdf}, based on 103 photometric measurements over the course of 27 years.
\subsubsection{Spectral energy distribution}\label{subsubsection_spectral_energy_distribution}
We model the spectral energy distribution ($F_{\lambda}=F_{\text{Bright/Faint}}\cdot f_{\alpha_{\text{Bright/Faint}}}$) of bright and faint stars in the FOV with a power law \citep{gravity_collaboration_deep_2022},
\begin{equation}
    f_{\alpha_{\text{Bright/Faint}}}\ =\ \left(\frac{\lambda}{\SI{2.2}{\mu\meter}}\right)^{-\alpha_{\text{Bright/Faint}}-1},
\end{equation}
approximating it with the Rayleigh-Jeans law \citep{buchholz_composition_2009}. We assume the spectral index to be normally distributed
\begin{equation}
    \alpha_{\text{Bright/Faint}}\ \sim\ \mathcal{N}\left(\mu_{\alpha\ {\text{Bright/Faint}}}, \sigma_{\alpha\ {\text{Bright/Faint}}}^2\right)
\end{equation}
with $\mu_{\alpha\ {\text{Bright/Faint}}}=\SI{3}{}$ and $\sigma_{\alpha\ {\text{Bright/Faint}}}=\SI{1.0}{}$.
\subsection{Faint stars}\label{subsection_faint_stars}
The faint model is the non-parameterized part of the object model; its purpose is to account for any faint, yet undiscovered sources in the FOV. We do not aim to model extended emission, and limit this model to stars that appear as unresolved point sources to GRAVITY. We effectively introduce a continuous background field consisting of pixels, and expect them to be uncorrelated.
\subsubsection{Flux density}
\begin{figure}[H]
    \centering
    \includegraphics[width=\linewidth]{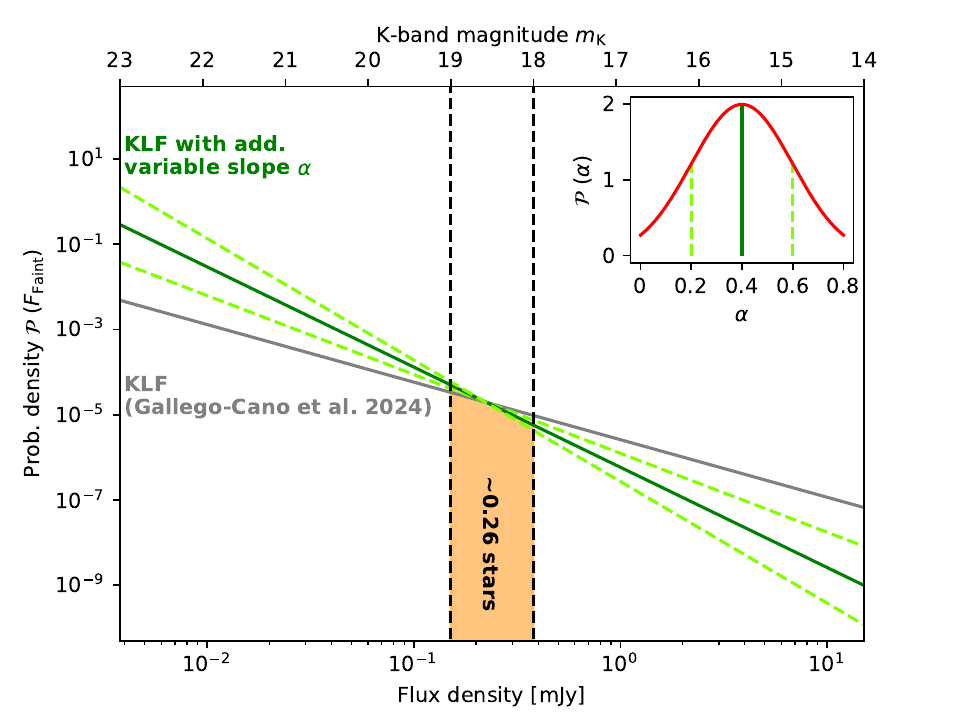}
    \caption{Faint flux density model in $\text{G}^{\text{R}}$, based on the KLF of early-type stars from literature \citep{gallego-cano_photometric_2024}, modified by an additional variable slope to allow for the inference of background flux, while the amount of expected stars in the FOV between $m_{\text{K}}=18$ and $m_{\text{K}}=19$ is kept constant at a literature value of $n_{\text{stars}}=\SI{0.26}{}$ stars \citep[derived from][]{gallego-cano_photometric_2024}.}
    \label{fig_faint_flux}
\end{figure}

Every pixel in the faint object model follows the same flux density prior, based on the K-band luminosity function (KLF), that is, the stellar surface density per K-band magnitude bin in the GC \citep{genzel_stellar_2003,buchholz_composition_2009,sabha_s-star_2012,schodel_milky_2020,gallego-cano_photometric_2024}. The KLF $\frac{dN\left(m_{\text{K}}\right)}{dm_{\text{K}}}$ in the GC is modeled by a linear slope over magnitudes, parameterized by the slope parameter $\beta$ \citep{genzel_stellar_2003},
\begin{equation}
    \frac{d}{dm_{\text{K}}}\text{log}\left(N\left(m_{\text{K}}\right)\right)\ =\ \beta.
\end{equation}
Converting this expression to flux densities, the corresponding PDF is proportional to $F_{\text{Faint}}^{-2.5\cdot\beta-1}$,
\begin{equation}
    \mathcal{P}\left(F_{\text{Faint}}\right)\propto\frac{dN\left(F_{\text{Faint}}\right)}{dF_{\text{Faint}}}\ =\ F_{\text{Faint}}^{-2.5\cdot\beta-1}\frac{2.5\cdot10^{m_{\text{K},\text{norm}}\cdot\beta}\cdot\beta\cdot C}{F_{\text{norm}}^{-2.5\cdot\beta}}.
\end{equation}
However, since GRAVITY has a finite limiting magnitude, an additional incoherent background flux is present. Hence we extend the KLF model by adding the parameter $\alpha$ to the slope parameter $\beta$,
\begin{equation}
    \mathcal{P}\left(F_{\text{Faint}}\right)\propto\frac{dN\left(F_{\text{Faint}}\right)}{dF_{\text{Faint}}}\propto F_{\text{Faint}}^{-2.5\cdot\left(\alpha+\beta\right)-1}.
\end{equation}
We deliberately renounce from modeling the incoherent flux density as part of the normalization in the response equation (cf. Eq. \ref{eq_response}) as all incoherent flux is absorbed in the $\alpha$ parameter and hence in the faint object model, and facilitates a statement on the significance of a source detection (\ref{subsection_statistical_significance}).\\
We obtain the PDF for the faint flux model by normalizing the previous expression
\begin{flalign}
    & \mathcal{P}\left(F_{\text{Faint}}\right)\ =\ \frac{\frac{dN\left(F_{\text{Faint}}\right)}{dF_{\text{Faint}}}}{\int_{F_{\text{limit}}}^{\infty} d{F_{\text{Faint}}}\frac{dN\left(F_{\text{Faint}}\right)}{dF_{\text{Faint}}}}\ =\ \\
    & 2.5\cdot\left(\alpha+\beta\right)\cdot F_{\text{Faint}}^{-2.5\cdot\left(\alpha+\beta\right)-1}\cdot F_{\text{limit}}^{2.5\cdot\left(\alpha+\beta\right)},
\end{flalign}
and introduce a lower flux density cutoff $F_{\text{limit}}$ to prevent divergence. While the $\alpha$ parameter in the model is flexible, and is inferred when $\text{G}^{\text{R}}$ approaches convergence (cf. Appendix \ref{appendix_inference}), we demand that the expected number of stars between $m_{\text{K}}=18$ and $m_{\text{K}}=19$ in the FOV remains fixed
\begin{equation}
    \int_{F\left(m_{\text{K}}=19\right)}^{F\left(m_{\text{K}}=18\right)} d{F_{\text{Faint}}}\ \mathcal{P}\left(F_{\text{Faint}}\right)\ \overset{!}{=}\ \frac{n_{\text{stars}}}{N_{\text{pixel}}},
\end{equation}
to the value from literature $n_{\text{stars}}=\SI{0.26}{}$ \citep[derived from][]{gallego-cano_photometric_2024}.\\
Solving this equation for $F_{\text{limit}}$, we obtain
\begin{equation}
    F_{\text{limit}}\ =\ \left(\frac{F\left(m_{\text{K}}=18\right)^{2.5\cdot\left(\alpha+\beta\right)}F\left(m_{\text{K}}=19\right)^{2.5\cdot\left(\alpha+\beta\right)}n_{\text{stars}}}{\left(F\left(m_{\text{K}}=18\right)^{2.5\cdot\left(\alpha+\beta\right)}-F\left(m_{\text{K}}=19\right)^{2.5\cdot\left(\alpha+\beta\right)}\right)N_{\text{pixel}}}\right)^{\frac{1}{2.5\cdot\left(\alpha+\beta\right)}},
\end{equation}
making $\mathcal{P}\left(F_{\text{Faint}}\right)$ solely dependent on $\alpha$, $\beta$, $n_{\text{stars}}$, and $N_{\text{pixel}}$.\\
We aim to infer the additional slope parameter $\alpha$ along with an image of the GC and hence derive a corresponding prior distribution. We demand that the flux density across the FOV integrated from $F_{\text{limit}}$ to $m_{\text{K}}=19$ to be equal to the faint flux present in the GC with the additional expected incoherent background flux, which we expect to be $F\left(m_{\text{K}, \text{limit}}\simeq16\right)$,
\begin{equation}
    N_{\text{pixel}}\int_{F_{\text{limit}}}^{F\left(m_{\text{K}}=19\right)} d{F_{\text{Faint}}}\ F_{\text{Faint}}\mathcal{P}\left(F_{\text{Faint}}\right)\ \overset{!}{=} F\left(m_{\text{K}, \text{limit}}\simeq16\right),
\end{equation}
corresponding to an additional slope parameter $\alpha\simeq\SI{0.4}{}$.\\
We thus impose a normal prior distribution on $\alpha$,
\begin{equation}
    \alpha\ \sim\ \mathcal{N}\left(\mu_{\alpha}, \sigma_{\alpha}^2\right)
\end{equation}
with a mean of $\mu_{\alpha}\simeq\SI{0.4}{}$ and a standard deviation of $\sigma_{\alpha}\simeq\SI{0.2}{}$.\\
The overall faint flux density model is illustrated in Fig. \ref{fig_faint_flux}.
\subsubsection{Spectral energy distribution}
The spectral index for faint stars is the same as for bright stars $\alpha_{\text{Faint}} = \alpha_{\text{Bright}}$, where a large width of the corresponding prior (\ref{subsubsection_spectral_energy_distribution}) accounts for a possible deviation from the Rayleigh-Jeans law for high magnitudes.
\subsection{Overall object model}\label{subsection_overall_object_model}
The previous aspects of the Galactic Center model taken together, the entire object model in $\text{G}^{\text{R}}$ is given by,
\begin{flalign}\label{eq_gc_model}
    & F\left(\boldsymbol{s},\lambda,t,p\right)\ =\ f_{\alpha_{\text{SgrA*}}}\left(\lambda\right)F_{\text{SgrA*}}\left(t,p\right)\delta\left(\boldsymbol{\boldsymbol{s}-\boldsymbol{s}_{\text{SgrA*}}}\right)+ \\
    & f_{\alpha_{\text{Bright, Faint}}}\left(\lambda\right)\left(\sum_{n=1}^{N_{\text{Bright}}}F_{\text{Bright, n}}\delta\left(\boldsymbol{\boldsymbol{s}-\boldsymbol{s}_{\text{Bright, n}}}\right)+F_{\text{Faint}}\left(\boldsymbol{s}\right)\right).
\end{flalign}
\section{Instrument response of GRAVITY}\label{appendix_response}
We augment the van Cittert-Zernike theorem (Eq. \ref{eq_van_cittert-zernike}) by systematic effects to arrive at a response equation for GRAVITY. These effects comprise optical aberrations along the GRAVITY beam train \citep{abuter_improved_2021}, as already implemented in \cite{gravity_collaboration_deep_2022}, and the bandpass across the finite spectral bandwidth \citep{widmann_improving_2018}.\\
We include the spectral bandpass in the response equation of GRAVITY, with the effect of refining the model towards the edge of the FOV, hence potentially increasing the sensitivity in a reconstructed image.
\subsection{Optical aberrations in GRAVITY}
In \cite{abuter_improved_2021}, the optical aberrations within the GRAVITY beam combiner instrument have been measured and subsequently fitted with a model for focal and pupil plane aberrations. The resulting fits are complex maps $A_{\text{j}}\left(\boldsymbol{s}\right)e^{i\phi_{\text{j}}\left(\boldsymbol{s}\right)}$ effectively encoding transmission and additional optical path difference (OPD) per telescope $\text{j}$, depending on the spatial vector $\boldsymbol{s}$.\\
The response equation of GRAVITY incorporates these aberrations in the form of photometric \citep{abuter_improved_2021},
\begin{equation}
    L_{\text{j}}\left(\boldsymbol{s}\right)\ =\ A_{\text{j}}\left(\boldsymbol{s}\right)^{2}
\end{equation}
and the interferometric lobes \citep{abuter_improved_2021}
\begin{equation}
    L_{\text{j,k}}\left(\boldsymbol{s}\right)\ =\ A_{\text{j}}\left(\boldsymbol{s}\right)e^{i\phi_{\text{j}}\left(\boldsymbol{s}\right)}A_{\text{k}}\left(\boldsymbol{s}\right)e^{-i\phi_{\text{k}}\left(\boldsymbol{s}\right)},
\end{equation}
for a baseline consisting of two telescopes $\text{j}$ and $\text{k}$.
\begin{figure*}
    \centering
    \includegraphics[width=\linewidth]{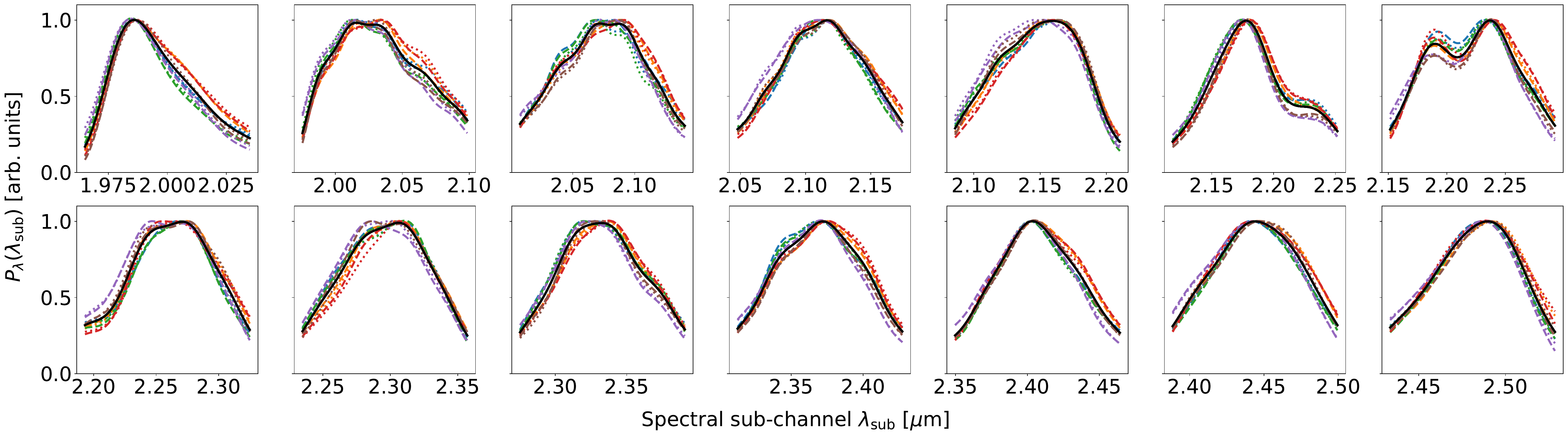}
    \caption{Spectral bandpass of GRAVITY in low spectral resolution, for six baselines (colors) and two linear polarizations (dashed/dotted). For every spectral channel, the mean over baselines and polarizations is plotted in black.}
    \label{fig_GRAVITY_bandpass}
\end{figure*}
These lobes are convolved with a normal distribution of $\sigma_{\text{t}}=\SI{10}{mas}$ standard deviation, to account for residual AO jitter \citep{abuter_improved_2021,perrin_single-mode_2019} during observations, which evaluates for the photometric lobes to
\begin{equation}
    \left\langle L_{\text{j}}\left(\boldsymbol{s}\right)\right\rangle_{\text{turb}}\ =\ A_{\text{j}}\left(\boldsymbol{s}\right)^{2}\circledast e^{-\frac{\boldsymbol{s}^2}{2\sigma_{\text{t}}^2}},
\end{equation}
and for the interferometric lobes to
\begin{equation}\label{eq_interferometric_lobes}
    \left\langle L_{\text{j,k}}\left(\boldsymbol{s}\right)\right\rangle_{\text{turb}}\ =\ \left(A_{\text{j}}\left(\boldsymbol{s}\right)e^{i\phi_{\text{j}}\left(\boldsymbol{s}\right)}\circledast e^{-\frac{\boldsymbol{s}^2}{2\sigma_{\text{t}}^2}}\right)^{*}\left(A_{\text{k}}\left(\boldsymbol{s}\right)e^{i\phi_{\text{k}}\left(\boldsymbol{s}\right)}\circledast e^{-\frac{\boldsymbol{s}^2}{2\sigma_{\text{t}}^2}}\right).
\end{equation}
The interferometric lobes effectively correct the coherent flux in the van Cittert-Zernike theorem, while the photometric lobes ensure proper normalization,
\begin{equation}\label{eq_reponse_aberr}
    V_{\text{j,k}}\left(\boldsymbol{u}\right)=\frac{\int\,d\boldsymbol{s}\ \left\langle L_{\text{j,k}}\left(\boldsymbol{s}\right)\right\rangle_{\text{turb}}F\left(\boldsymbol{s}\right)\ e^{-2\pi i\boldsymbol{s}\frac{\boldsymbol{B}_{\text{j,k}}}{\lambda}}}{\sqrt{\int\,d\boldsymbol{s}\ \left\langle L_{\text{j}}\left(\boldsymbol{s}\right)\right\rangle_{\text{turb}}F\left(\boldsymbol{s}\right)\int\,d\boldsymbol{s}\ \left\langle L_{\text{k}}\left(\boldsymbol{s}\right)\right\rangle_{\text{turb}}F\left(\boldsymbol{s}\right)}}.
\end{equation}
\subsection{Spectral bandpass in GRAVITY}
We account for the finite spectral bandwidth in GRAVITY by dividing every spectral channel $\lambda$ into 100 sub-channels $\lambda_{\text{sub}}$, and subsequently integrating Eq. \ref{eq_reponse_aberr} with the spectral bandpass in place. Contrary to \cite{gravity_collaboration_deep_2022}, we don't use a top-hat bandpass but instead the measured bandpass of GRAVITY $P_{\lambda}\left(\lambda_{\text{sub}}\right)$ in low spectral resolution
\begin{equation}\label{eq_response}
    V_{\text{j,k}}\left(\boldsymbol{u}\right)=\frac{\int\,d\lambda_{\text{sub}}\ P_{\lambda}\left(\lambda_{\text{sub}}\right)\int\,d\boldsymbol{s}\ \left\langle L_{\text{j,k}}\left(\boldsymbol{s}\right)\right\rangle_{\text{turb}}F\left(\boldsymbol{s}\right)\ e^{-2\pi i\boldsymbol{s}\frac{\boldsymbol{B}_{\text{j,k}}}{\lambda_{\text{sub}}}}}{\Pi_{\text{x}=\text{j,k}}\sqrt{\int\,d\lambda_{\text{sub}}\ P_{\lambda}\left(\lambda_{\text{sub}}\right)\int\,d\boldsymbol{s}\ \left\langle L_{\text{x}}\left(\boldsymbol{s}\right)\right\rangle_{\text{turb}}F\left(\boldsymbol{s}\right)}}.
\end{equation}
To keep the formalism clear to follow, we abbreviate the actual implemented spectral bandpass $P_{\lambda\ |\ \text{j,k}}\left(\lambda_{\text{sub}}, p\right)$ (cf. Fig. \ref{fig_GRAVITY_bandpass}), which is dependent on baseline $\text{j,k}$ and polarization $p$ to $P_{\lambda}\left(\lambda_{\text{sub}}\right)$, and further use $F\left(\boldsymbol{s}\right)$ for $F\left(\boldsymbol{s},\lambda\right)$, and $\left\langle L_{\text{x}}\left(\boldsymbol{s}\right)\right\rangle_{\text{turb}}$ for $\left\langle L_{\text{x}}\left(\boldsymbol{s},\lambda\right)\right\rangle_{\text{turb}}$.
\subsection{Self-calibration of visibility phases}
The concept of self-calibration stems from radio interferometry \citep{readhead_mapping_1978}, in which the missing visibility phase is reconstructed from closure phases and visibility amplitudes. GRAVITY, on the other hand, does measure visibility phases, but we aim to follow along this approach to infer minor inconsistencies in data from the data itself. Small variations in observing conditions impact observables, and we alleviate these effects by applying a time-dependent calibrations for both visibility phases and amplitudes \citep{gravity_collaboration_deep_2022}.\\
To calibrate visibility phases, we model fluctuations in OPD per baseline $C_{\phi\ |\ \text{j,k}}\left(t_{\text{m}}\right)$ and further parameterize this correction by introducing a time-dependent phase calibration $C_{\phi\ |\ \text{j}}\left(t_{\text{m}}\right)$ per telescope,
\begin{equation}\label{eq_visphi_selfcal}
    C_{\phi\ |\ \text{j,k}}\left(t_{\text{m}}\right)\ =\ \text{exp}\left(\frac{2\pi i}{\lambda}\left(C_{\phi\ |\ \text{j}}\left(t_{\text{m}}\right)-C_{\phi\ |\ \text{k}}\left(t_{\text{m}}\right)\right)\right).
\end{equation}
We assume $C_{\phi\ |\ \text{j}}\left(t_{\text{m}}\right)$ to follow a normal distribution,
\begin{equation}
    C_{\phi\ |\ \text{j}}\left(t_{\text{m}}\right)\ \sim\ \mathcal{N}\left(\mu_{\phi}, \sigma_{\phi}^2\right),
\end{equation}
with $\mu_{\phi}=0$ and constrain it with $\sigma_{\phi}=\SI{44}{\nano\meter}$, corresponding to the 50th part of $\SI{2.2}{\micro\meter}$. Visibility phases are only sensitive to the relative phase between telescopes of a corresponding baseline, and we can thus deliberately fix the phase calibration of one of four telescopes to zero.\\
We apply $C_{\phi\ |\ \text{j,k}}\left(t_{\text{m}}\right)$ to the exponential in Eq. \ref{eq_response} to arrive at phase calibrated visibilities.
\subsection{Self-calibration of visibility amplitudes}
Similar to the self-calibration of visibility phases, we introduce a calibration factor $C_{\rho\ |\ \text{j,k}}\left(t_{\text{m}}\right)$ for visibility amplitudes to absorb time-dependent variations in visibility amplitudes. This factor is further parameterized by a correction affecting the baseline $C_{0}\left(t_{\text{m}}\right)$ and a correction per telescope $C_{\text{j/k}}\left(t_{\text{m}}\right)$,
\begin{equation}
    C_{\rho\ |\ \text{j,k}}\left(t_{\text{m}}\right)\ =\ C_{\rho\ |\ 0}\left(t_{\text{m}}\right)+C_{\rho\ |\ \text{j}}\left(t_{\text{m}}\right)\cdot C_{\rho\ |\ \text{k}}\left(t_{\text{m}}\right).
\end{equation}
This approach allows us to decouple contributions that may arise from an elevated incoherent flux level and photometric fluctuations on individual telescopes.\\
We assign both $C_{0}\left(t_{\text{m}}\right)$ and $C_{\text{j/k}}\left(t_{\text{m}}\right)$ normal distributions
\begin{equation}
    C_{\rho\ |\ 0}\left(t_{\text{m}}\right)\ \sim\ \mathcal{N}\left(\mu_{\rho\ |\ 0}, \sigma_{\rho\ |\ 0}^2\right),
\end{equation}
with $\mu_{\rho,0}=0$, $\sigma_{\rho,0}=4$ and
\begin{equation}
    C_{\rho\ |\ \text{j}}\left(t_{\text{m}}\right)\ \sim\ \mathcal{N}\left(\mu_{\rho\ |\ \text{j}}, \sigma_{\rho\ |\ \text{j}}^2\right),
\end{equation}
with $\mu_{\rho}=1$ and a constraining $\sigma_{\rho}=0.05$. Analogously to the self-calibration approach for visibility phases, we keep $C_{\text{j/k}}\left(t_{\text{m}}\right)$ fixed for one telescope.\\
The calibration factor is multiplied to Eq. \ref{eq_response} to obtain visibilities with calibrated amplitudes,
\begin{equation}\label{eq_visamp_selfcal}
    V_{\text{j,k}}^{\text{cal}}\left(\boldsymbol{u}\right)=V_{\text{j,k}}\left(\boldsymbol{u}\right)\ C_{\rho\ |\ \text{j,k}}\left(t_{\text{m}}\right).
\end{equation}
We clip the samples of $C_{\rho\ |\ 0}\left(t_{\text{m}}\right)$ to zero to prevent that negative samples induce negative visibility amplitudes.
\subsection{Reduced number of FFT evaluations for GRAVITY+}\label{subsection_reduced_number_of_FFT}
The extreme AO of GRAVITY+ injects more light into the fibers of the GRAVITY beam combiner instrument, allowing for GC observations in medium spectral resolution mode (233 spectral channels). The evaluation of the forward model in $\text{G}^{\text{R}}$ becomes very time intensive as the interferometric lobe in the numerator of the response equation (Eq. \ref{eq_response}) is in fact dependent on the observed wavelength, $\left\langle L_{\text{j,k}}\right\rangle_{\text{turb}}\left[\boldsymbol{s}, \lambda\right]$, that was omitted for brevity. This results in the evaluation of $n_{\text{baselines}}\ \times\ n_{\text{ch}}\ =\ 1398$ Fast Fourier Transforms (FFTs), which are dominating by far the total runtime of the reconstruction method. In principle, there is no way to circumvent this tedious computation; however, we can exploit the dependency on wavelength of the optical aberrations in GRAVITY. The aberration map of a spectral channel $\lambda_{1}$, can be approximated by the map of another spectral channel $\lambda_{2}$, scaled by the factor $\frac{\lambda_{2}}{\lambda_{1}}$ \citep{abuter_improved_2021},
\begin{equation}
    A_{\text{j}}\left(\boldsymbol{s},\lambda_{1}\right)e^{i\phi_{\text{j}}\left(\boldsymbol{s},\lambda_{1}\right)}\ \simeq\ A_{\text{j}}\left(\boldsymbol{s}\frac{\lambda_{2}}{\lambda_{1}},\lambda_{2}\right)e^{i\phi_{\text{j}}\left(\boldsymbol{s}\frac{\lambda_{2}}{\lambda_{1}},\lambda_{2}\right)}.
\end{equation}
We can thus approximate the convoluted interferometric lobe with
\begin{equation}
    \left\langle L_{\text{j,k}}\right\rangle_{\text{turb}}\left[\boldsymbol{s},\lambda_{1}\right]\ \simeq \ \left(\frac{\lambda_{1}}{\lambda_{2}}\right)^2\left\langle L_{\text{j,k}}\right\rangle_{\text{turb}}\left[\boldsymbol{s}\frac{\lambda_{2}}{\lambda_{1}},\lambda_{2}\right],
\end{equation}
where we effectively absorb $\frac{\lambda_{2}}{\lambda_{1}}$ into the convolution kernel $\sigma_{\text{t}}$ (cf. Eq. \ref{eq_interferometric_lobes}), as its value of $\sigma_{\text{t}}\ =\ \SI{10}{mas}$ can vary anyway \citep{perrin_single-mode_2019,abuter_improved_2021}.\\
The goal is to compute the response equation of GRAVITY, without changing the integrand. The numerator in Eq. \ref{eq_reponse_aberr} can then be expressed as
\begin{flalign}
    & \int\,d\boldsymbol{s}\ \left\langle L_{\text{j,k}}\right\rangle_{\text{turb}}\left[\boldsymbol{s},\lambda_{1}\right]\ F\left(\boldsymbol{s}\right)\ e^{-2\pi i\boldsymbol{s}\frac{\boldsymbol{B}_{\text{j,k}}}{\lambda_{2}}}\ =\ \\
    & \int\,d\boldsymbol{s}\ \left\langle L_{\text{j,k}}\right\rangle_{\text{turb}}\left[\boldsymbol{s},\lambda_{1}\right]\ e^{-2\pi i\boldsymbol{s}\frac{\boldsymbol{B}_{\text{j,k}}}{\lambda_{2}}}\circledast\int\,d\boldsymbol{s}\ F\left(\boldsymbol{s}\right)\ e^{-2\pi i\boldsymbol{s}\frac{\boldsymbol{B}_{\text{j,k}}}{\lambda_{2}}}\ =\ \\
    & \left(\frac{\lambda_{1}}{\lambda_{2}}\right)^2\int\,d\boldsymbol{s}\ \left\langle L_{\text{j,k}}\right\rangle_{\text{turb}}\left[\boldsymbol{s}\frac{\lambda_{2}}{\lambda_{1}},\lambda_{2}\right]\ e^{-2\pi i\boldsymbol{s}\frac{\boldsymbol{B}_{\text{j,k}}}{\lambda_{2}}}\circledast\\
    & \int\,d\boldsymbol{s}\ F\left(\boldsymbol{s}\right)\ e^{-2\pi i\boldsymbol{s}\frac{\boldsymbol{B}_{\text{j,k}}}{\lambda_{2}}}\ =\ \\
    & \left(\frac{\lambda_{1}}{\lambda_{2}}\right)^4\int\,d\boldsymbol{s}\ \left\langle L_{\text{j,k}}\right\rangle_{\text{turb}}\left[\boldsymbol{s},\lambda_{2}\right]\ e^{-2\pi i\boldsymbol{s}\frac{\boldsymbol{B}_{\text{j,k}}}{\lambda_{2}}\left(\frac{\lambda_{1}}{\lambda_{2}}\right)}\circledast\\
    & \int\,d\boldsymbol{s}\ F\left(\boldsymbol{s}\right)\ e^{-2\pi i\boldsymbol{s}\frac{\boldsymbol{B}_{\text{j,k}}}{\lambda_{2}}},\\
\end{flalign}
where $\circledast$ denotes convolution. In principle, the exponential $e^{-2\pi i\boldsymbol{s}\frac{\boldsymbol{B}_{\text{j,k}}}{\lambda_{2}}\left(\frac{\lambda_{1}}{\lambda_{2}}\right)}$ can be expanded as a Taylor series for $\left(\frac{\lambda_{1}}{\lambda_{2}}\right)$ around 1, but any corrections arising from this expansion are leading to complicated expressions that would defy a simplification of the forward model. We thus take only to the zeroth order into account and introduce the approximation $e^{-2\pi i\boldsymbol{s}\frac{\boldsymbol{B}_{\text{j,k}}}{\lambda_{2}}\left(\frac{\lambda_{1}}{\lambda_{2}}\right)}\ \simeq\ e^{-2\pi i\boldsymbol{s}\frac{\boldsymbol{B}_{\text{j,k}}}{\lambda_{2}}}$.\\
Taken together, we can approximate visibilities $V_{\text{j,k}}$ as visibilities with a wavelength independent integrand $V^{\text{indp}}_{\text{j,k}}$ by applying a correction factor,
\begin{equation}
    V_{\text{j,k}}\left(\frac{\boldsymbol{B}_{\text{j,k}}}{\lambda_{2}}\right)\ \simeq\ \left(\frac{\lambda_{1}}{\lambda_{2}}\right)^4\cdot V^{\text{indp}}_{\text{j,k}}\left(\frac{\boldsymbol{B}_{\text{j,k}}}{\lambda_{2}}\right).
\end{equation}
This allows us to compute only a single FFT per baseline in the forward model and makes $\text{G}^{\text{R}}$ applicable to datasets acquired in medium and, in principle, even high spectral resolution. To keep errors as small as possible, we choose one of the central spectral channels of GRAVITY as $\lambda_{1}$.
\begin{figure*}[t!]
    \centering
    \includegraphics[width=\linewidth]{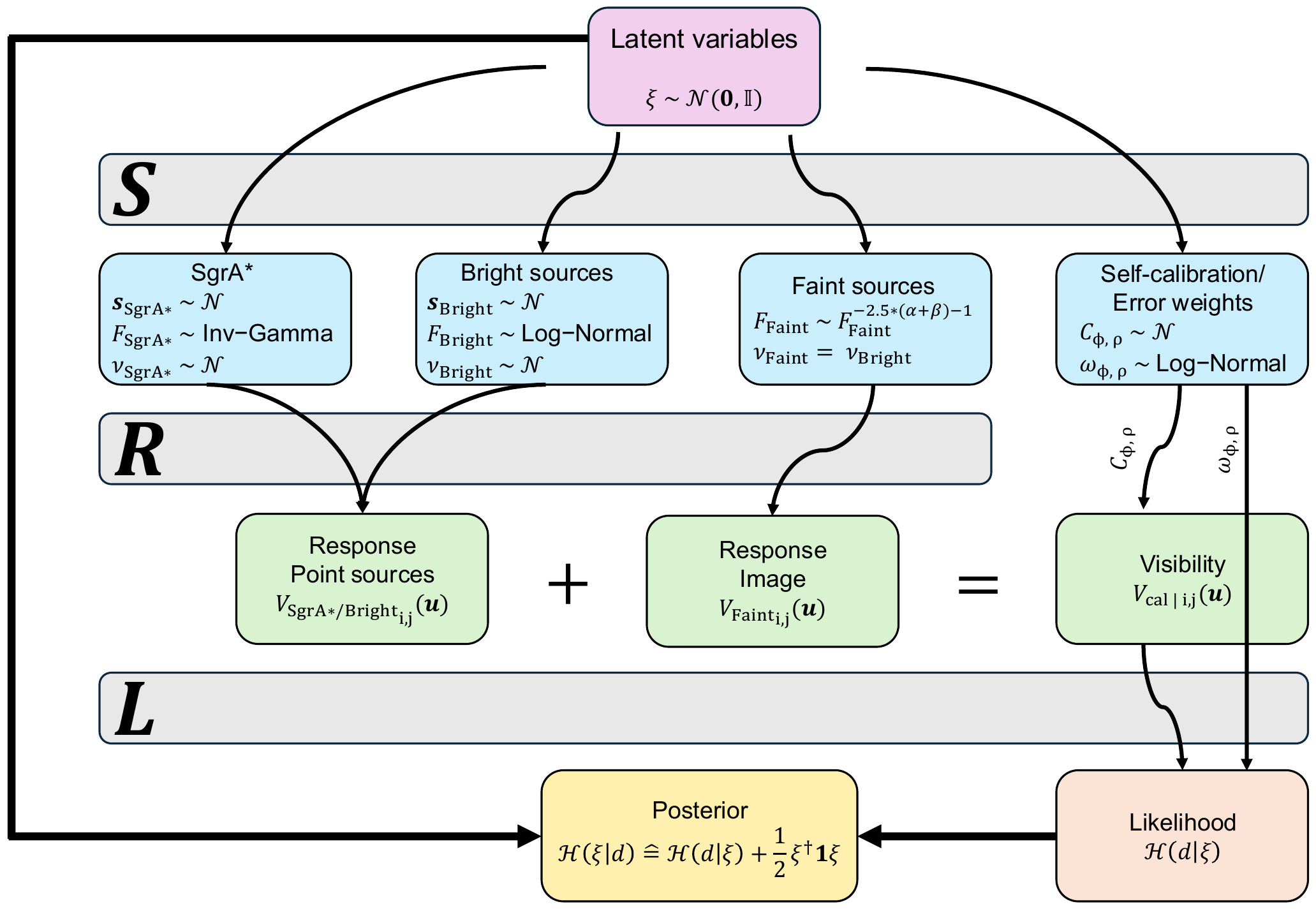}
    \caption{Schematic overview on the forward model in $\text{G}^{\text{R}}$. Coloured boxes represent different hierarchies within the model, from lowest (top) to highest (bottom). Grey boxes depict the object $\mathcal{S}$, response $\mathcal{R}$, and likelihood $\mathcal{L}$ operators, respectively. Black arrows crossing grey boxes indicate the action of the respective operator on the quantity of the arrow's origin, with the computed result at their tip. The basis of the forward model is the set of latent variables $\xi_{\text{i}}$ that is mapped by $\mathcal{S}$ to parameters of the object model, self-calibration, and error weights model. By acting $\mathcal{R}$ on the object model, one obtains the response model, while the application of $\mathcal{L}$ yields the likelihood.\\
    In the case of mosaicing, this model is modified as outlined in Fig. \ref{fig_Mosaic_model}.}
    \label{fig_hierarchical_model}
\end{figure*}
\section{Likelihood}\label{appendix_likelihood}
Several processes contribute to noise in GRAVITY data, and hence we assume, in line with the central limit theorem, that uncertainties are normally distributed. The likelihood probability density distribution (PDF) in $\text{G}^{\text{R}}$ is hence normal distributed. We treat visibility phases and visibility amplitudes separately, and construct covariance matrices for both data products
\begin{equation}\label{eq_cov_phi}
    \Sigma_{\phi}\ =\ \sigma_{\phi\ |\ \text{j,k}}\left(\lambda\right)^{2}\mathbf{1},
\end{equation}
\begin{equation}\label{eq_cov_rho}
    \Sigma_{\rho}\ =\ \sigma_{\rho\ |\ \text{j,k}}\left(\lambda\right)^{2}\mathbf{1},
\end{equation}
respectively by using the $\SI{1}{}\sigma$ error per datapoint as provided by the GRAVITY pipeline \citep{lapeyrere_gravity_2014}.\\
To account for wrapping of the visibility phase, we convert the data phases $\phi_{\text{data}}$ and model $\phi_{\text{data}}$ to complex numbers on the unit circle in the complex plane, with Eq. \ref{eq_cov_phi} being the corresponding covariance matrix,
\begin{flalign}
    & \mathcal{P}\left(d-\mathcal{R}\left(F\right)|F\right)_{\phi}\ =\ \mathcal{P}\left(\phi_{\text{data}}-\phi_{\text{model}}\right)\ \simeq\ \\
    & \mathcal{N}\left(e^{i\phi_{\text{data}}}-e^{i\phi_{\text{model}}},\sigma_{\phi\ |\ \text{j,k}}\left(\lambda\right)^{2}\mathbf{1}\right).
\end{flalign}
The likelihood PDF for visibility amplitudes with the diagonal covariance matrix of Eq. \ref{eq_cov_rho}, we arrive correspondingly at
\begin{flalign}
    & \mathcal{P}\left(d-\mathcal{R}\left(F\right)|F\right)_{\rho}\ =\ \mathcal{P}\left(\rho_{\text{data}}-\rho_{\text{model}}\right)\ =\ \\
    & \mathcal{N}\left(\rho_{\text{data}}-\rho_{\text{model}},\sigma_{\rho\ |\ \text{j,k}}\left(\lambda\right)^{2}\mathbf{1}\right).
\end{flalign}
By applying the information hamiltonian $\mathcal{H}\left(x\right)\ =\ -\text{log}\left(\mathcal{P}\left(x\right)\right)$ \citep{enslin_information_2019,knollmuller_metric_2020} we can add up both contributions and obtain, up to constant factors $\left(\hat{=}\right)$, 
\begin{flalign}
    & \mathcal{H}\left(d|F\right)\ =\ \mathcal{H}\left(d|F\right)_{\phi}+\mathcal{H}\left(d|F\right)_{\rho}\ \hat{=}\ \\
    & \frac{1}{2}\left(e^{i\phi_{\text{data}}}-e^{i\phi_{\text{model}}}\right)^{\dagger}\Sigma_{\phi}^{-1}\left(e^{i\phi_{\text{data}}}-e^{i\phi_{\text{model}}}\right)+\\
    & \frac{1}{2}\left(\rho_{\text{data}}-\rho_{\text{model}}\right)^{\dagger}\Sigma_{\rho}^{-1}\left(\rho_{\text{data}}-\rho_{\text{model}}\right).
\end{flalign}
This represents the quantity, along with the prior, that is minimized in MGVI \citep{knollmuller_metric_2020}.
\subsection{Weighted error scaling}
Error quantification can be subject to uncertainty itself, resulting in a potentially underestimated noise level. From experience, the errors as computed by the GRAVITY pipeline \citep{lapeyrere_gravity_2014} are notoriously underestimated by mostly a factor of $\sim 1.45$. We aim to take this into account and use the concept of dynamic error scaling \citep{arras_comparison_2021} and infer a correction factor to errors along with the image itself.\\
We hence apply a correction factor $w_{\phi}$ to the diagonal entries of the visibility phase covariance matrix
\begin{equation}
    \Sigma_{\phi}\ =\ \left(w_{\phi}\ \sigma_{\phi\ |\ \text{j,k}}\left(\lambda\right)\right)^{2}\mathbf{1},
\end{equation}
and correspondingly apply the factor $w_{\rho}$ to the visibility amplitude covariance matrix
\begin{equation}
    \Sigma_{\rho}\ =\ \left(w_{\rho}\ \sigma_{\rho\ |\ \text{j,k}}\left(\lambda\right)\right)^{2}\mathbf{1}.
\end{equation}
Both weighting factors follow a log-normal prior distribution, imposing a mode on $\SI{1.45}{}$ and a median at $\SI{2.0}{}$,
\begin{equation}
    w_{\phi,\rho}\ \sim\ \text{LogNormal}\left(\mu=0.693,\sigma=0.567\right).
\end{equation}
Additionally, we clip this distribution to a lower value of 1.0 to safeguard against any weighting below the error estimate of the GRAVITY pipeline.
\begin{figure*}[t!]
    \centering
    \includegraphics[width=\linewidth]{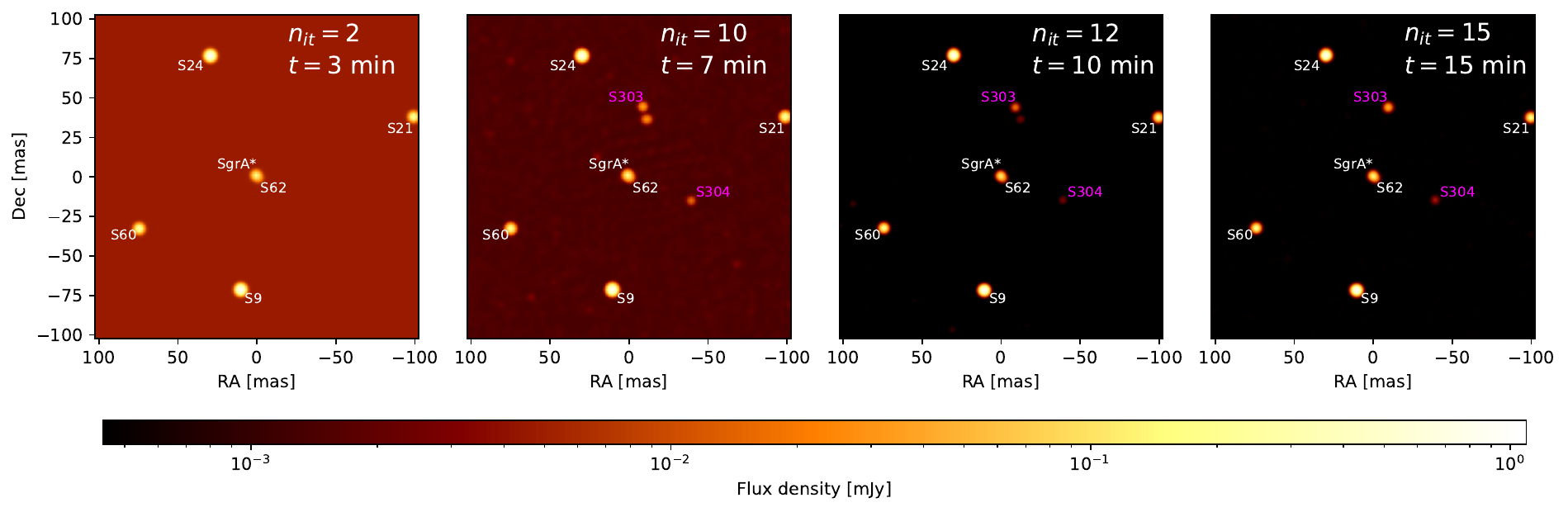}
    \caption{Evolution of the reconstructed image during inference. Between iterations $n_{\text{it}}=10$ and $n_{\text{it}}=12$, $\text{G}^{\text{R}}$ is refining the image by quadrupling the number of pixels.}
    \label{fig_GPAO_image_inference}
\end{figure*}
\begin{table*}[t!]
    \centering
    \resizebox{\linewidth}{!}{
    \begin{tabular}{c|c|c|c|c|c|c|c|c|c|c|c!{\vrule width 3pt}c|c|c|c|c}
         & \multicolumn{3}{c|}{$n\left(\text{d. o. f.}\right) < n\left(\text{data}\right)$} & \multicolumn{13}{c}{$n\left(\text{d. o. f.}\right) > n\left(\text{data}\right)$} \\
        $n_{\text{it}}$ & 0 & 1 & 2 & 3 & 4 & 5 & 6 & 7 & 8 & 9 & 10 & 11 & 12 & 13 & 14 & 15\\\hline\hline
        $F_{\text{Faint}}$ & \cellcolor{red!50} Fix &  \cellcolor{red!50} Fix & \cellcolor{red!50} Fix & \cellcolor{green!50} Free & \cellcolor{green!50} Free & \cellcolor{green!50} Free & \cellcolor{green!50} Free & \cellcolor{green!50} Free & \cellcolor{green!50} Free & \cellcolor{green!50} Free & \cellcolor{green!50} Free & \cellcolor{green!50} Free & \cellcolor{green!50} Free & \cellcolor{green!50} Free & \cellcolor{green!50} Free & \cellcolor{green!50} Free \\\hline
        $\alpha$ & \cellcolor{green!50} Free &  \cellcolor{red!50} Fix & \cellcolor{red!50} Fix  & \cellcolor{red!50} Fix & \cellcolor{red!50} Fix & \cellcolor{red!50} Fix & \cellcolor{red!50} Fix & \cellcolor{red!50} Fix & \cellcolor{red!50} Fix & \cellcolor{red!50} Fix & \cellcolor{green!50} Free & \cellcolor{green!50} Free & \cellcolor{green!50} Free & \cellcolor{green!50} Free & \cellcolor{green!50} Free & \cellcolor{green!50} Free \\\hline
        $w_{\phi,\rho}$ & \cellcolor{green!50} Free &  \cellcolor{green!50} Free & \cellcolor{green!50} Free  & \cellcolor{red!50} Fix & \cellcolor{red!50} Fix & \cellcolor{red!50} Fix & \cellcolor{red!50} Fix & \cellcolor{red!50} Fix & \cellcolor{red!50} Fix & \cellcolor{red!50} Fix & \cellcolor{red!50} Fix & \cellcolor{red!50} Fix & \cellcolor{red!50} Fix & \cellcolor{red!50} Fix & \cellcolor{red!50} Fix & \cellcolor{red!50} Fix \\\hline
        $C_{\phi\ |\ \text{j}}\left(t_{\text{m}}\right)$ & \cellcolor{green!50} Free &  \cellcolor{green!50} Free & \cellcolor{green!50} Free  & \cellcolor{red!50} Fix & \cellcolor{red!50} Fix & \cellcolor{red!50} Fix & \cellcolor{red!50} Fix & \cellcolor{red!50} Fix & \cellcolor{red!50} Fix & \cellcolor{red!50} Fix & \cellcolor{red!50} Fix & \cellcolor{red!50} Fix & \cellcolor{red!50} Fix & \cellcolor{red!50} Fix & \cellcolor{red!50} Fix & \cellcolor{red!50} Fix \\\hline
        $C_{\rho\ |\ \text{j}}\left(t_{\text{m}}\right)$ & \cellcolor{green!50} Free &  \cellcolor{green!50} Free & \cellcolor{green!50} Free  & \cellcolor{red!50} Fix & \cellcolor{red!50} Fix & \cellcolor{red!50} Fix & \cellcolor{red!50} Fix & \cellcolor{red!50} Fix & \cellcolor{red!50} Fix & \cellcolor{red!50} Fix & \cellcolor{red!50} Fix & \cellcolor{red!50} Fix & \cellcolor{red!50} Fix & \cellcolor{red!50} Fix & \cellcolor{red!50} Fix & \cellcolor{red!50} Fix \\\hline
        $C_{\rho\ |\ 0}\left(t_{\text{m}}\right)$ & \cellcolor{red!50} Fix & \cellcolor{green!50} Free & \cellcolor{green!50} Free & \cellcolor{red!50} Fix & \cellcolor{red!50} Fix & \cellcolor{red!50} Fix & \cellcolor{red!50} Fix & \cellcolor{red!50} Fix & \cellcolor{red!50} Fix & \cellcolor{green!50} Free & \cellcolor{red!50} Fix & \cellcolor{red!50} Fix & \cellcolor{red!50} Fix & \cellcolor{red!50} Fix & \cellcolor{red!50} Fix & \cellcolor{red!50} Fix \\\hline
        $F_{\text{SgrA*}}\left(t_{\text{m}}\right)$ & \cellcolor{green!50} Free & \cellcolor{green!50} Free & \cellcolor{green!50} Free & \cellcolor{red!50} Fix & \cellcolor{red!50} Fix & \cellcolor{red!50} Fix & \cellcolor{red!50} Fix & \cellcolor{red!50} Fix & \cellcolor{red!50} Fix & \cellcolor{red!50} Fix & \cellcolor{red!50} Fix & \cellcolor{red!50} Fix & \cellcolor{red!50} Fix & \cellcolor{red!50} Fix & \cellcolor{red!50} Fix & \cellcolor{green!50} Free \\\hline
        Samples (KL) & 12 & 12 & 12 & 12 & 12 & 12 & 12 & 12 & 12 & 12 & 24 & 24 & 24 & 24 & 24 & 24 \\\hline
        Iteration limit (KL) & 250 & 250 & 250 & 250 & 250 & 250 & 250 & 250 & 250 & 250 & 250 & 250 & 250 & 250 & 250 & 250 \\\hline
    \end{tabular}
    }
    \caption{Inference scheme as applied in $\text{G}^{\text{R}}$, see subsection \ref{subsec_inference_strategy} for details. The thick vertical line marks the transition from a low-resolution to a high-resolution image.}
    \label{tab_inference_scheme}
\end{table*}
\section{Inference}\label{appendix_inference}
We apply the Bayesian inference method Metric Gaussian Variational Inference (MGVI) \citep{knollmuller_metric_2020}, to explore the very high-dimensional posterior of the problem (subsection \ref{subsection_inference}). Following subsection \ref{subsection_likelihood} we write the corresponding posterior as
\begin{equation}\label{eq_bayes_theorem_F}
    \mathcal{P}\left(F|d\right)\ \propto\ \mathcal{P}\left(d-\mathcal{R}\left(F\right)|F\right)\cdot\mathcal{P}\left(F\right),
\end{equation}
being solely dependent on the object model $F$. The individual components of the object model follow different prior distributions (see section \ref{appendix_gc_model}, Fig. \ref{fig_hierarchical_model}), and we reduce the complexity of Eq. \ref{eq_bayes_theorem_F} by introducing a set of standardized coordinates, which allows us to apply MGVI.

\subsection{Standardized coordinate system of variational parameters}\label{subsection_standardization}
As part of the forward model, we map every parameter of the object model $F\left(\boldsymbol{s}_{\text{SgrA*}}, F_{\text{SgrA*}}, \nu_{\text{SgrA*}}, \boldsymbol{s}_{\text{Bright}}, ...\right)$ to corresponding standardized coordinates $F\left(\boldsymbol{\xi}_{\boldsymbol{s}_{\text{SgrA*}}}, \xi_{F_{\text{SgrA*}}}, \xi_{\nu_{\text{SgrA*}}}, \boldsymbol{\xi}_{\boldsymbol{s}_{\text{Bright}}}, ...\right)$ that are all standard normally distributed. We reduce the hierarchical structure of the forward model with the 'parametrization trick' \citep{kingma_auto-encoding_2022,knollmuller_metric_2020}, which in practice amounts to inverse transform sampling \citep{devroye_sample-based_1986},
\begin{equation}
    f\left(\xi\right)\ =\ \left(F_{\mathcal{P}}^{-1}\circ F_{\mathcal{N}\left(0,1\right)}\right)\left[\xi\right],
\end{equation}
where $F$ denotes the cumulative distribution function (CDF) for the corresponding distribution in its subscript.\\
The application of the object model operator $\mathcal{S}\left[\boldsymbol{\xi}\right]$ in Eq. \ref{eq_sky_op} equals the mapping $f\left(\xi\right)$.\\
This allows us to write the information Hamiltonian in a much simpler form,
\begin{equation}
        \mathcal{H}\left(\xi|d\right)\ \hat{=}\ \mathcal{H}\left(d|\xi\right)+\frac{1}{2}\xi^{\dagger}\mathbf{1}\xi,
\end{equation}
and represents the posterior MGVI is exploring.\\
Finally, the reconstructed image and its corresponding statistics are all encoded in the $\boldsymbol{\xi}$ and can be obtained by the application of the object model operator $\mathcal{S}$. Equivalently, to retrieve the response model and its statistics, as well as possibly correlations, the chained operator of response and object operator $\mathcal{R}\circ\mathcal{S}$ is applied.
\begin{figure}[H]
    \centering
    \includegraphics[width=\linewidth]{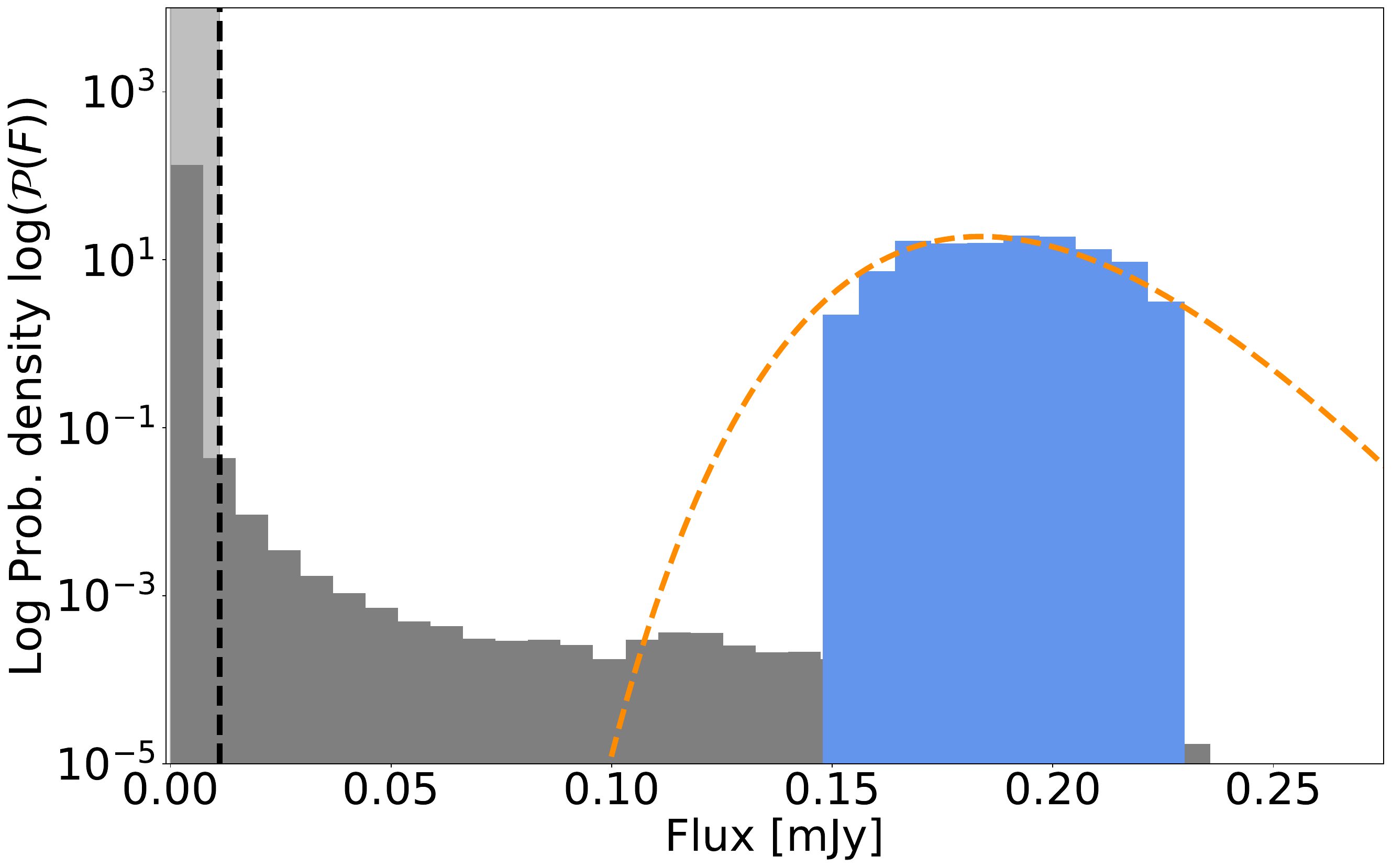}
    \caption{Flux density samples of an injected source (blue histogram) and the faint object model (grey histogram). The statistical significance of the injected source against background is determined by discarding samples beyond $m_{\text{K}}=22$, fitting a log-normal distribution (orange line) to the remaining samples, and applying Eq. \ref{eq_statistical_significance}.}
    \label{fig_detection_significance_new}
\end{figure}
\subsection{Inference strategy}\label{subsec_inference_strategy}
The inference in $\text{G}^{\text{R}}$ is focused on robustness of results, deep images, while providing computational speed. This is achieved by removing degeneracies, increasingly complicating the forward model following \citep{edenhofer_sparse_2022}, and choosing educated hyperparameters for MGVI \citep{knollmuller_metric_2020}.\\
The response equation of GRAVITY (Eq. \ref{eq_response}) is scale invariant with respect to the object model $F\left(\boldsymbol{s}\right)$, and we deliberately fix the flux density of the brightest source in the FOV. The underlying assumption is that its photometry is known well enough that it can serve as a reference to other, potentially faint sources within the image. Moreover, we fix the self-calibration parameters for visibility phase and amplitude, $C^{\ \phi}_{\text{j}}\left(t_{\text{m}}\right)$ and $C^{\ \rho}_{\text{j}}\left(t_{\text{m}}\right)$, for one telescope $\text{j}$ without constraining the model.\\
The inference scheme in $\text{G}^{\text{R}}$ follows the strategy to provide first the simplest model without major constraints and further complicate it step by step (cf. Tab. \ref{tab_inference_scheme}). In the first three global iterations of MGVI \citep{knollmuller_metric_2020}, we do not take individual pixels of the faint object model into account, which leads to a strongly over-determined inference problem. MGVI is highly efficient in this regime, and we use this setting to infer the slope of the faint object model, all self-calibration parameters, and the lightcurve of SgrA*. The remaining iteration steps are focused on inferring faint sources, which is alleviated by temporarily increasing the slope of the faint object model to facilitate sampling with MGVI. This results in a temporarily higher background level, which is absorbed by the $C^{\ \rho}_{0}\left(t_{\text{m}}\right)$ parameter.\\
The set of hyperparameters of MGVI that have the most impact on the performance of $\text{G}^{\text{R}}$ are the number of samples and the number of iterations that are used to minimize the Kullback-Leibler divergence \citep{kullback_information_1951} (hereafter KL-divergence) with the Vector-free L-BFGS optimizer \citep{chen_large-scale_2014}, which is available in NIFTy. By choosing more samples and more iterations, the forward model in $\text{G}^{\text{R}}$ is evaluated more often, which ultimately leads to a slower inference. In our hyperparameter setting, we try to ensure robustness and precision of results, while saving computational cost. Throughout the inference, we select a high iteration limit for the optimizer (Tab. \ref{tab_inference_scheme}) to ensure highest precision that is computationally feasible and only vary the number of samples. We start the inference at a high number of samples and only switch towards the end of the inference, when the object model has switched to a higher spatial resolution, to the highest number of samples that are within our computational budget. MGVI allows us to parallelize the computation of samples, and we run $\text{G}^{\text{R}}$ typically on 24 CPUs, which makes 24 samples the final sensible number of samples in the inference. From experience, choosing either too few samples or iterations to compute the KL-divergence has a drastic effect on statistical robustness, that is, images computed with different random seeds start to vary in appearance up to the point that spurious sources appear. As soon as $\text{G}^{\text{R}}$ starts to converge, typically the case after 11 global iterations, the inference is carried out in a high-resolution grid with four times as many pixels.
\subsection{Statistical significance}\label{subsection_statistical_significance}
The detection significance of injected sources (Fig. \ref{fig_injection_map}, \ref{subsubsection_limiting_K_band_magnitude}) is determined from the flux density distributions of the injected source and the faint object model (Fig. \ref{fig_detection_significance_new}). The flux density histogram of the injected source is truncated to $m_{\text{K}}=\SI{22}{}$, and a log-normal distribution is fitted to the remaining samples. We arrive at a significance estimate by determining the median of the faint object model samples, subtracting it from the median of the log-normal fit, and by dividing the result by the square root of the log-normal variance. Finally, we weight the result by the number of flux density samples of the injected source brighter than $m_{\text{K}}=\SI{22}{}$ divided by the total number of samples,
\begin{equation}\label{eq_statistical_significance}
    \sigma_{\text{signf}}\ =\ \frac{\text{exp}\left(\mu_{\text{inj}}\right)-\Tilde{F}_{\text{faint}}}{\sqrt{\left(\text{exp}\left(\sigma_{\text{inj}}^2\right)-1\right)\text{exp}\left(2\mu_{\text{inj}}+\sigma_{\text{inj}}^2\right)}}\cdot\frac{n_{\text{inj}}\left(m_{\text{K}}<22\right)}{n_{\text{inj}}}.
\end{equation}
\end{appendix}

\end{document}